\documentclass[usenatbib]{mnras}

\usepackage{newtxmath}

\usepackage[T1]{fontenc}
\usepackage{ae,aecompl}

\usepackage{graphicx}	
\usepackage{amsmath}	
\usepackage{mathrsfs}	

\usepackage{ulem}
\usepackage{longtable}
\usepackage{color}
\usepackage{xspace}
\usepackage{subcaption}
\graphicspath{{figures/}}

\newcommand{\Msun}{\ensuremath{\mathrm{M}_\odot}} 

\newcommand{\ergs}{\ensuremath{\mathrm{erg\,s}^{-1}}}
\newcommand{\s}{{\ensuremath{\mathrm{s}}}}
\newcommand{\kms}{{\ensuremath{\mathrm{km}\,\s^{-1}}}}

\newcommand{\modelza}{\texttt{z25a}\xspace}
\newcommand{\modelzb}{\texttt{z25b}\xspace}
\newcommand{\modeloc}{\texttt{35OC}\xspace}
\newcommand{\modelzal}{\texttt{z25a-01}\xspace}
\newcommand{\modelzbl}{\texttt{z25b-01}\xspace}
\newcommand{\modelocl}{\texttt{35OC-01}\xspace}
\newcommand{\modelzam}{\texttt{z25a-05}\xspace}
\newcommand{\modelzbm}{\texttt{z25b-05}\xspace}
\newcommand{\modelocm}{\texttt{35OC-05}\xspace}
\newcommand{\modelzah}{\texttt{z25a-10}\xspace}
\newcommand{\modelzbh}{\texttt{z25b-10}\xspace}
\newcommand{\modeloch}{\texttt{35OC-10}\xspace}

\title[Jet-Driven Hypernovae]{The chemical signature of jet-driven hypernovae}

\author[J.J.\ Grimmett et al.]{
J.J. Grimmett,$^{1,5}$\thanks{E-mail: james.grimmett@monash.edu}
Bernhard M\"uller,$^{1-3}$
Alexander Heger,$^{1,3-6}$
\newauthor
Projjwal Banerjee,$^{7}$
\& Martin Obergaulinger$^{8,9}$
\\
$^{1}$School of Physics and Astronomy, 19 Rainforest Walk, Monash University, VIC 3800, Australia\\
$^{2}$Astrophysics Research Centre, School of Mathematics and Physics, Queen's University Belfast, BT7 1NN, Belfast, Northern Ireland\\
$^{3}$Australian Research Council Centre of Excellence for Gravitational Wave Discovery (OzGrav), Clayton, VIC 3800, Australia\\
$^{4}$Center of Excellence for Astrophysics in Three Dimensions (ASTRO-3D), Australia\\
$^{5}$Joint Institute for Nuclear Astrophysics, 1 Cyclotron Laboratory, National Superconducting Cyclotron Laboratory,\\ \phantom{$^5$}Michigan State University, East Lansing, MI 48824-1321, USA\\
$^{6}$Tsung-Dao Lee Institute, Shanghai 200240, China\\
$^{7}$Discipline of Physics, Indian Institute of Technology Palakkad, Palakkad, Kerala, India 678557\\
$^{8}$Departament d'Astonomia i Astrof\'isca, Universitat de Val\`encia, Edifici d'Investigatci\'o Jeroni Munyoz, C/Dr.\ Moliner, 50,\\
\phantom{$^8$}E-46100 Burjassot (Val\`encia),
Spain
\\
$^{9}$Institut f\"ur Kernphysik, Technische Universit\"at Darmstadt, Schlossgartenstr. 2, 64289 Darmstadt, Germany
}

\date{Accepted XXX. Received YYY; in original form ZZZ}

\pubyear{xxxx}

\begin{document}

\label{firstpage}
\pagerange{\pageref{firstpage}--\pageref{lastpage}}
\maketitle
 
\begin{abstract}
Hypernovae powered by magnetic jets launched from the surface of rapidly rotating millisecond magnetars are one of the leading models to explain broad-lined Type Ic supernovae (SNe Ic-BL), and have been implicated as an important source of metal enrichment in the early Universe.  We investigate the nucleosynthesis in such jet-driven hypernovae using a parameterised, but physically motivated, approach that analytically relates an artificially injected jet energy flux to the power available from the energy in differential rotation in the proto-neutron star. We find ejected $^{56}\mathrm{Ni}$ masses of $0.05\,\Msun \texttt{- }0.45\,\Msun$ in our most energetic models with explosion energy $>10^{52}\,\mathrm{erg}$.  This is in good agreement with the range of observationally inferred values for SNe Ic-BL.  The $^{56}\mathrm{Ni}$ is mostly synthesised in the shocked stellar envelope, and is therefore only moderately sensitive to the jet composition. Jets with a high electron fraction $Y_\mathrm{e}=0.5$ eject more $^{56}\mathrm{Ni}$ by a factor of 2 than neutron-rich jets. We can obtain chemical abundance profiles in good agreement with the average chemical signature observed in extremely metal-poor (EMP) stars presumably polluted by hypernova ejecta.  Notably, $\mathrm{[Zn/Fe]} \gtrsim 0.5$ is consistently produced in our models.  For neutron-rich jets, there is a significant \textsl{r}-process component, and agreement with EMP star abundances in fact requires either a limited contribution from neutron-rich jets or a stronger dilution of \textsl{r}-process material in the interstellar medium than for the slow SN ejecta outside the jet.  The high $\mathrm{[C/Fe]}\gtrsim 0.7$ observed in many EMP stars cannot be consistently achieved due to the large mass of iron in the ejecta, however, and remains a challenge for jet-driven hypernovae based on the magneto-rotational mechanism.
\end{abstract}




\section{Introduction}\label{sec:intro}

A new observational class of extreme supernovae
emerged in the late 1990s with the observation of the unusually bright supernova (SN) 1998bw and
the accompanying long gamma-ray burst (GRB) 980425. Several spectroscopically similar SNe, both with and without GRBs,
have been observed since SN 1998bw \citep[see, e.g.,][]{mazzali_2014, anderson_2019, taddia_2019}.
One-dimensional supernova models suggested that
an explosion energy of order $10^{52}\,\mathrm{erg}$, and more than $0.1\,\Msun$ of $^{56}\mathrm{Ni}$ in the ejecta 
are required to reproduce the bright peak in optical luminosity and the high ejecta velocities inferred
from the broad spectral lines of these events
(e.g., SN 1998bw, $20 \texttt{-} 50 \times 10^{51}\,\mathrm{erg}$, \citealt{iwamoto_1998,nakamura_2001a}; SN 2003dh, $26\times10^{51}\,\mathrm{erg}$, \citealt{woosley_2003}). 
Such large values of explosion energy and $^{56}$Ni  mass are each an order of magnitude more than the values inferred
from observational properties of more commonly observed core-collapse supernovae.

The bright luminosity and large explosion energy has earned these SNe the colloquial name 'hypernovae', though the exact requirements to earn the classification are loosely defined. Often the name is used to designate bright supernovae with an estimated explosion energy of at least $10\times10^{51}\,\mathrm{erg}$, other times a companion GRB seems to be an additional requirement. Sometimes the term
is also extended to the more narrowly defined class of superluminous supernovae, though
it is usually applied to a specific sub-class
of stripped-envelope supernovae
characterised by broad spectral lines and the absence of hydrogen and helium features, the broad-lined Type Ic (Ic-BL) supernovae .
For clarity, we will henceforth refer to the observed transients as SNe Ic-BL, and to hypernovae as the physical model(s) proposed to explain the observations. 

Both SNe Ic-BL and long GRBs are preferentially observed in low-metallicity galaxies at high redshift \citep{gehrels_2009,arcavi_2010,smith_2011}. 
Additionally, observations of GRBs indicate that they tend to be located in the star-forming regions of their host galaxies \citep{paczynski_1998,woosley_2006}. 
The preference for these environment suggests
an explosion mechanism that is related to peculiarities
of massive star evolution at low-metallicity, namely reduced
mass and angular momentum losses.
It is unclear, however, what the precise evolutionary pathways of hypernova progenitors are, and how they are able to release such a large amount of energy in their explosions \citep[see, e.g.,][]{yoon_05,yoon_2006,woosley_2006a,detmers_08,woosley_2011,aguilera_18,aguilera_20}.

The observational features of Ic-BL SNe suggest a prominent
role of jets --  in the broad sense of bipolar outflows -- and point away from the neutrino-driven mechanism. The latter is believed to power the majority of ordinary supernova explosions, but
is limited to energies $\mathord{\lesssim} 2 \times 10^{51}\, \mathrm{erg}$ \citep{thielemann_1996,muller_2016,muller_2017,janka_2017,burrows_2020}.
Spectroscopic and spectropolarimetric observations of SNe Ic-BL provide consistent evidence for a nearly axisymmetric bipolar explosion geometry, which points away from models based on the canonical neutrino-driven mechanism \citep{wang_2008,stevance_2017,ashall_2019}.
The features of the typical Ic-BL light curve and optical spectra both indicate that at least two distinct components exist in the ejecta, providing the signature of asphericity. One component is rapidly expanding and iron-rich, whereas the other is slowly expanding and oxygen-rich \citep{mazzali_2001,maeda_2008,tanaka_2017}. This, in addition to the GRB connection, points toward jet-driven explosions, the idea being that a strong polar shock will result in a rapidly expanding and iron-rich jet-like component of the ejecta, whereas the weakly shocked equatorial material provides the slowly expanding oxygen-dominated component. The relativistic  GRB jet must be a distinct phenomenon, though possibly related to the non-relativistic bipolar
outflows that carry the bulk of the explosion energy.

Such bipolar explosions could most naturally be explained by rotation and magnetic fields in the progenitor. The models implicated to explain the Ic-BL observational signature, along with the occasional companion GRB, are collectively referred to as hypernovae. The two most widely discussed engine models are the collapsar model, and the magneto-rotational model, which is also known as the millisecond magnetar model. 

The collapsar model involves the ongoing evolution of "failed" supernovae during the collapse of massive, rotating stars, wherein the large gravitational potential of a massive core inhibits a successful neutrino-driven explosion either by 
thwarting shock revival altogether or by massive fallback \citep{macfadyen_1999,macfadyen_2001}. Instead, the core collapses to a black hole (BH) surrounded by a
a rotationally-supported accretion disk
from which large amounts of rotational energy may be extracted
by magnetic fields, and possibly by neutrino-antineutrino
annihilation, to power jet ouflows into the polar direction
as well as also powerful disk winds.

An alternative scenario involves magneto-rotational explosions
feeding on the rotational energy of a rapidly spinning (period $P\sim1\,\mathrm{ms}$) ``millisecond magnetar'': Once sufficiently strong fields are generated by rotational winding, the magneto-rotational instability (MRI;
\citealp{akiyama_2003,thompson_2005}) and/or dynamo amplification \citep{mosta_2015, raynaud_2020},
MHD jets are launched as a result of extreme magnetic pressure at the poles and collimated by hoop stresses \citep{burrows_2007,mosta_2014,obergaulinger_2020}. During the early non-relativistic
phase of jet propagation, the jets impart energy to the stellar envelope, driving an explosion and limiting continued accretion onto the core. 

There is some circumstantial evidence in favour of
the millisecond magnetar scenario. The rotational energy of a rapidly rotating protoneutron star ($10^{52}\,\mathrm{erg}$) available to power MHD jets is remarkably similar to the explosion energy inferred in several observed SNe Ic-BL \citep{mazzali_2014}. Plateaus in the X-ray afterglows
of long GRBs possibly indicate the presence of a magnetar
\citep{corsi_09,gompertz_14}, though alternative explanations exist \citep{duffell_15}.
Finally, the shape of some hypernova light curves have sometimes been explained by energy input from magnetar spin-down
\citep[e.g.,][]{woosley_2010,greiner_2015,wang_2017,wang_2019}.
Many uncertainties remain about the magneto-rotational mechanism, however, such as the precise workings of the various field amplification mechanisms and issues such as jet stability \citep{mosta_2014,kuroda_2020}.

Regardless of the hypernova mechanism, there is evidence for a significant role of hypernovae in the chemical evolution of
the early Universe.
Many long-lived extremely metal-poor (EMP) stars that have presumably formed from gas polluted by no more than a few
supernovae exhibit abundance patterns that cannot be explained from the yields of core-collapse supernovae of typical explosion energy $\sim 10^{51}\,\mathrm{erg}$ \citep{mcwilliam_1995a,mcwilliam_1995b,ryan_1996,cayrel_2004}. 
Based on artificial 1D  models for hypernova explosions
it has been suggested
that their characteristic nucleosynthesis, along with the large mass of $^{56}$Ni needed to power the HN light curve, could provide a better explanation, e.g., for the supersolar [(Co,Zn)/Fe] observed in EMP stars
\citep[e.g.,][]{nakamura_2001,umeda_2002,kobayashi_2006}
as well as the high [C/Fe] in some of them \citep{tominaga_2009,ezzeddine_2019}.  There
are also arguments that hypernovae may be needed as an additional
source for rapid-neutron capture process (\textsl{r}-process) elements at low metallicity to supplement the nucleosynthethis contribution
of neutron star mergers \citep{winteler_2012,nishimura_2017,mosta_2018,kobayashi_2020}.

Evidently, multi-dimensional simulations are required to properly
model the explosion dynamics and the nucleosynthesis conditions in jet-driven explosions, however.  By now, a large number of 
simulations have already comprehensively explored the nucleosynthesis in collapsars using parameterised two-dimensional (2D) hydro simulations with
prescribed outflow boundary conditions for the jets \citep{maeda_2002,maeda_2003,nagataki_2006,tominaga_2009,barnes_2018}, or with more consistent 2D MHD
simulations \citep{ono_2009,ono_2012,fujimoto_2007,nakamura_2015}, which, however still cannot capture essentially
three-dimensional (3D) dynamo processes. The nucleosynthesis in magneto-rotational explosions powered by millisecond magnetars
has been studied both with 2D \citep{nishimura_2006,nishimura_2015,nishimura_2017} and 3D MHD simulations \citep{winteler_2012,moesta_2018,halevi_2018} with a view to the role of hypernovae as \textsl{r}-process production sites. These MHD models of magneto-rotational supernovae, however, have so far been restricted in terms of simulation time and hence may give a rather incomplete picture of the nucleosynthesis, though
\citet{reichert_20} have recently presented
nucleosynthesis from long-time MHD simulations
including neutrino transport in 2D.
In particular, the mass of $^{56}\mathrm{Ni}$ that can be synthesised in the millisecond-magnetar scenario has not yet been conclusively determined, though arguments have been made based on analytic estimates \citep{suwa_2015}.  First parameterised hydrodynamic simulations rather call into doubt whether sufficient $^{56}$Ni can be produced to match the high values derived from typical HN light curves \citep{chen_2017}. Though late-time powering by the magnetar wind
could also  explain key features of HN light curves
\citep[e.g.,][]{woosley_2010,chen_2017}, a more comprehensive study
of the nucleosynthesis in magneorotational explosions from
the light elements through $^{56}$Ni as key observable to
the neutron-rich heavy elements is desirable.

In order to obtain such comprehensive nucleosynthesis results from 
long-time simulations of jet-driven explosions based on the magneto-rotational mechanism, we here
adopt a parameterised approach with artificially injected
jets, similar to what has been done in the context of the collapsar \citep{maeda_2002,maeda_2003,nagataki_2006,tominaga_2009}. Even as first nucleosynthesis results from long-time 2D MHD simulations are becoming
available \citep{reichert_20}, such a parameterised
approach remains useful because it still allows
considerably longer simulation times, and since
even 2D MHD simulations remain subject to
significant uncertainties, e.g., the question
of non-axisymmetric effects on the propagation
of the jet and uncertainties in the jet
electron fraction, which is very sensitive
to details of the neutrino transport.
To adopt this approach to the millisecond magnetar scenario,
we use a different physical model for the jet based on the
notion that during the first seconds of the explosion
the dominant energy source for the jets is the free energy in the differential rotation \citep{burrows_2007}. Using this model, which
links the jet outflows to the proto-neutron star angular
momentum and the mass and angular momentum accretion rate,
we simulate explosions driven by artificially injected jets for three rapidly rotating
hypernova progenitor models. 
After describing our methodology in Section~\ref{sec:method},
we discuss the dynamics of our parameterised jet models,
the production of $^{56}\mathrm{Ni}$, and the synthesis
of other heavy elements against the backdrop of observed
abundances in EMP stars in Section \ref{sec:results}. We discuss the implicatons
of our finding for the viability of jet-driven explosions
in the millisecond magnetar scenario as an explanation
for Ic-BL SNe and EMP abundances in Section~\ref{sec:discussion}
and conclude with a short summary in Section~\ref{sec:summary}.


\section{Method}\label{sec:method}
\subsection{Progenitor models}\label{ssec:prog_models}

\begin{figure}
	\includegraphics[width=\columnwidth]{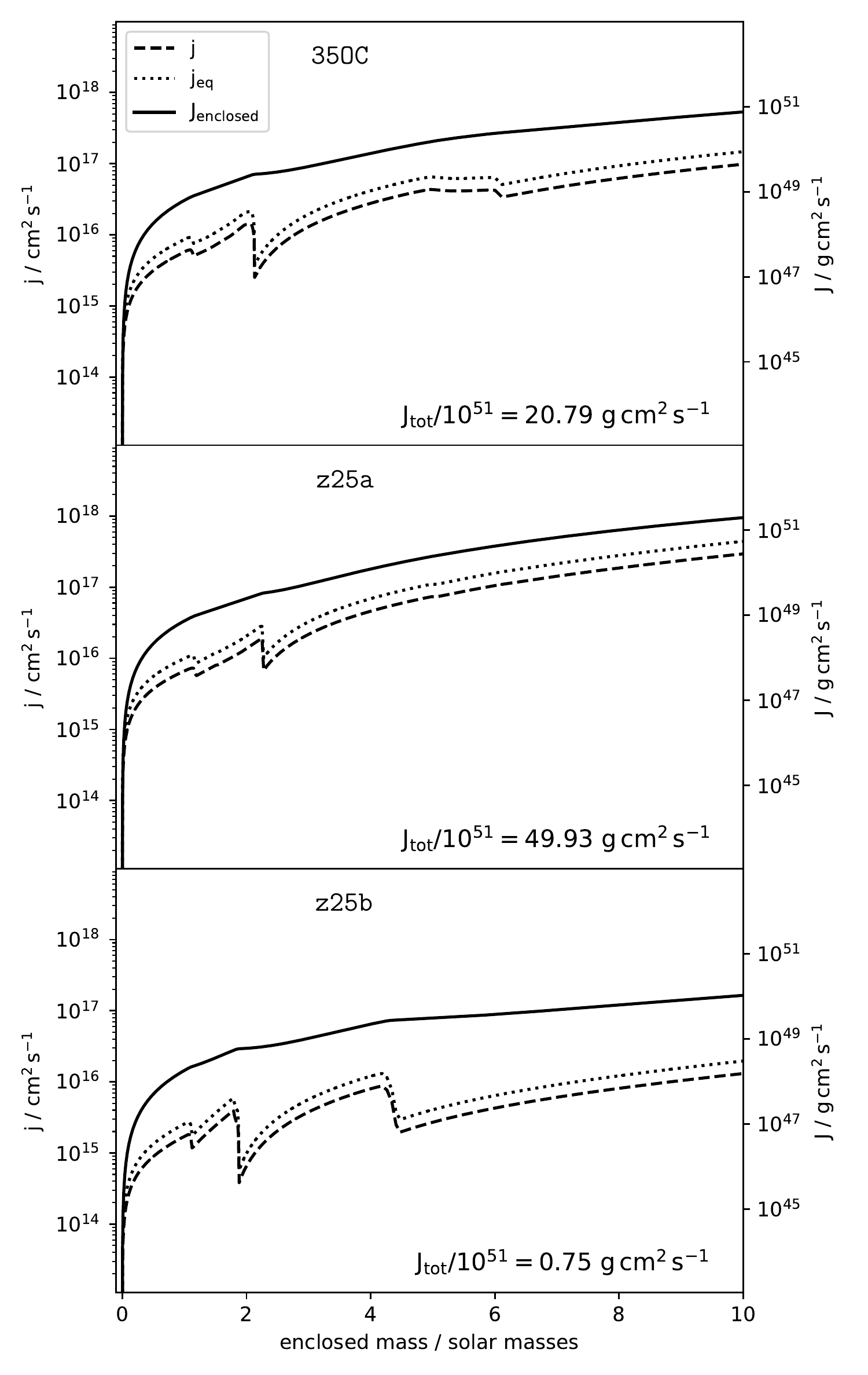}
	\caption{The shell-averaged (dashed line) and equatorial (dotted line) specific angular momentum profile for each progenitor model is shown on the left axes.  The right axes show the enclosed total angular momentum (solid line) for each progenitor model. The profiles are shown at the end of silicon burning, i.e., at the pre-supernova stage.}
	\label{fig:j_profiles}
\end{figure}

As progenitors, we use three massive low- or zero-metallicity rapidly-rotating stellar models, evolved to the onset of core-collapse. They include Model \texttt{35OC} from \citet{woosley_2006a} which has an  initial mass of $35\,\Msun$ with $10\,\%$ of solar metallicity, and two
zero-metallicity (Population III) Models \texttt{z25a} and \texttt{z25b} with an initial mass of $25\,\Msun$.  These models have been evolved using the stellar evolution code
\textsc{Kepler} \citep{weaver_1978}. Approximations for multi-dimensional effects due to rotation such as angular momentum transport and mixing are the same as presented in \citet{heger_2000}. The evolution of magnetic field resulting from Taylor-Spruit dynamo and the associated magnetic torque \citep{spruit_2002} is taken in to account as detailed in \citet{heger_2005}. All models have an initial rotation speed of $\sim 50\%$ of the critical breakup speed that corresponds to an equatorial surface rotation velocity of $\sim 380\,\kms$ for Model \texttt{35OC} and $\sim 700\, \kms$ for Models \modelza and \modelzb halfway through H burning. 

The mass loss rate from \citet{nieu_1990} is adopted for stars in the main sequence and red supergiant stages. A star is assumed be in the Wolf-Rayet (WR) stage when the effective surface temperature exceeds $10^4$~K and surface H mass fraction drops below $0.4$. The mass loss rate during the WR stage is adopted from \citet{hamann_1995} but lowered by a factor of 3 to account for clumping \citep{hamann_1998}. The mass loss rate is assumed to scale as $(Z/\mathrm{Z}_\odot)^{0.5}$ where $Z$ is the initial metallicity. Whereas Model \modelza does not experience any mass loss (since $Z=0$), Model \modelzb has been modified to shed mass at it would if it had $\mathrm{Z} = 10^{-3}\,\mathrm{Z}_\odot$ similar to the models presented in \citet{banerjee_2019}.
The relevant pre-collapse properties of each star are given in Table~\ref{table:progenitor_models}, and the angular momentum profiles are shown in Figure \ref{fig:j_profiles}.

\begin{table}
\caption{Properties of each model at the onset of collapse.}
\label{table:progenitor_models}
\begin{tabular}{ |c|c|c|c|c|c| } 
 \hline
\multicolumn{6}{c}{Presupernova models}\\
 \hline
 \hline
Model & Z & Mass & Radius & M$_\mathrm{Fe\,core}$ & J$_\mathrm{Fe\,core}$  \\
      & Z$\odot$ & \Msun & $10^{11}$ cm & \Msun & $10^{49}\,\mathrm{erg\,s}$ \\
\hline
    \texttt{35OC} & 0.1 & 28.07 & 1.56 & 2.02 & 2.30  \\
    \texttt{z25a} & 0.0 & 25.00 & 7.71 & 2.25 & 3.29  \\
    \texttt{z25b} & 0.0 & 20.09 & 5.37 & 1.79 & 0.54   \\
 \hline
\end{tabular}\\
\end{table}

\subsection{Simulations of Jet-Driven Explosions}
\label{sec:coconut}

We use the \textsc{CoCoNuT} code, a Godunov-based Eulerian  relativistic hydrodynamic solver \citep{dimmelmeier_2002,mueller_2010,mueller_2015} with higher-order reconstruction to follow the collapse and post-bounce evolution of these stars, after mapping each model to two dimensions. We assume equatorial and axial symmetry, and our computational domain is covered by 550 radial and 128 angular zones, extending out to a maximum radius of $2\times10^{10}\,\mathrm{cm}$. Until
$\mathord{\sim}80\,\mathrm{ms}$, the 
simulations include neutrino transport using the fast multi-group (FMT) scheme
of \citet{mueller_2015}. In the
high-density regime, we use the nuclear
equation of state of \citet{lattimer_1991} with an
incompressibility modulus of $K=220 \, \mathrm{MeV}$. At low densities, we treat the gas as en ensemble of nuclei, 
electrons, positrons, and photons. At
temperatures greater than $8\, \mathrm{GK}$,
nuclear statistical equilibrium is assumed.
At lower temperatures we use a
19-species network including
protons, neutrons,
${}^3\mathrm{He}$,
${}^{4}\mathrm{He}$
${}^{12}\mathrm{C}$,
${}^{14}\mathrm{N}$
${}^{16}\mathrm{O}$,
${}^{20}\mathrm{Ne}$,
${}^{24}\mathrm{Mg}$,
${}^{28}\mathrm{Si}$,
${}^{32}\mathrm{S}$,
${}^{36}\mathrm{Ar}$,
${}^{40}\mathrm{Ca}$,
${}^{44}\mathrm{Ti}$,
${}^{48}\mathrm{Cr}$,
${}^{52}\mathrm{Fe}$,
${}^{54}\mathrm{Fe}$, and
${}^{56}\mathrm{Ni}$
\citep{weaver_1978} to treat nuclear burning
and recombination\footnote{Additionally to the listed species, protons from photo-disintegration are treated separately as 19$^\mathrm{th}$ species.}.  In the NSE regime,
we additionally formally include a few
neutron-rich species 
(${}^{56}\mathrm{Fe}$, ${}^{60}\mathrm{Fe}$, ${}^{70}\mathrm{Ni}$
and very neutron-rich dummy species
${}^{120}\mathrm{Ni}$, $^{200}\mathrm{Zr}$) 
that can be formed during freeze-out from NSE at a low electron fraction $Y_\mathrm{e}$. In practice, neutron-rich
material enters the NSE regime in a dissociated
state dominated by $\alpha$-particles and free neutrons,
and due to the use of the 19-species network at
lower temperatures, we underestimate the recombination
to nuclei at low $Y_\mathrm{e}$ in the jet during
the hydro simulation. This does not significantly
affect the energetics of the explosion for several
reasons, however. The energetically more important process of recombination into $\alpha$-particles \emph{is}
treated accurately; recombination into nuclei is
incomplete in the jets because of their high entropies.

At $\sim 80\,\mathrm{ms}$ post-bounce, we excise the region inside a  radius $200\,\mathrm{km}$ and implement
bipolar jet outflows by prescribing
appropriate boundary conditions near the pole, as described in the following section. Outflow boundary conditions are used at lower latitudes. Neutrino transport is switched off, and the metric is frozen at this point.

\subsection{Prescription for jet energy}\label{ssec:jet_method}
Numerical MHD studies have shown that the magneto-rotational explosion mechanism can power bipolar jets in collapsing stars, as long as the magnetic field can continue to tap energy from the differentially rotating core \citep[e.g.,][]{akiyama_2003,blackman_2006,burrows_2007,obergaulinger_2017,obergaulinger_2018}.
Based on their models, \citet{burrows_2007} argued
that in a quasi-steady state with sufficiently strong magnetic fields, the power of the
magnetically-driven outflows is regulated by the rate
at which the accretion flow brings in additional
energy into differential rotation. Based on the
notion that the rate of increase of the free rotational
energy $\dot{E}_\mathrm{free}$ is balanced
by the jet power, we construct a simple
analytic model to relate the jet power to
the properties of the accretion flow at the
excision boundary.

In our model, we assume that the accreted
material is initially accreted onto and mixed
homogeneously into the proto-neutron star (PNS) without being braked into co-rotation. The rate
$\dot{E}_\mathrm{rot,acc}$ at which rotational energy is thus injected into the PNS by accretion is
\begin{equation}
\label{eq:erot1}
    \dot{E}_\mathrm{rot,acc}=\dot{M} \frac{j^2}{4 / 5\, R^2},
\end{equation}
where $\dot{M}$ is the mass accretion rate, $j$
is the specific angular momentum of the accreted matter, and $R$ is the PNS radius.
For simplicity, we assume that the PNS is a homogeneous sphere of radius $15\, \mathrm{km}$ to compute its radius of gyration. Deviations from
this (crude) assumption can be absorbed into an
overall efficiency factor for the jet power.

$\dot{E}_\mathrm{rot,acc}$ can then be compared to 
the rate of increase of the PNS rotational energy
after the accreted material has come into corotation
with the PNS, which is assumed to rotate uniformly.
Using a PNS moment of inertia $I=2 / 5\, M R^2$
in terms of PNS mass $M$, we obtain
\begin{align}
\nonumber
    \dot{E}_\mathrm{rot,PNS}&=\frac{\mathrm{d}}{\mathrm{d}t}\left(\frac{J^2}{2I}\right)=\frac{\mathrm{d}}{\mathrm{d}t}\left(\frac{J^2}{4 / 5 M R^2}\right)=
    \frac{5}{4 R^2}\left(\frac{2 \dot{J} J}{M}-
    \frac{\dot {M}J^2}{M^2}\right)\\
    &=\frac{5\left(2 \dot M j j_\mathrm{PNS}-\dot M j_\mathrm{PNS}^2\right)}{4 R^2}
    =\frac{5 \dot{M} \left(2 j j_\mathrm{PNS}- j_\mathrm{PNS}^2\right)}{4 R^2},
\end{align}
where $j_\mathrm{PNS}=J/M$ is the average specific angular momentum of the PNS. By subtracting 
$\dot{E}_\mathrm{rot,PNS}$ from $\dot{E}_\mathrm{rot,acc}$, we find that accretion
provides free rotational energy at a rate of
\begin{equation}
    \dot{E}_\mathrm{free}=
    \frac{5 \dot{M} \left(j^2-2 j j_\mathrm{PNS}+ j_\mathrm{PNS}^2\right)}{4 R^2}=
    \frac{5 \dot{M} \left(j-j_\mathrm{PNS}\right)^2}{4 R^2}.
\end{equation}
We assume that $\dot{E}_\mathrm{free}$ is converted
into jet power
$\dot{E}_\mathrm{jet}$ with an efficiency parameter,
$\epsilon$,
\begin{equation}\label{eq:e_jet}
	\dot{E}_\mathrm{jet} = \epsilon \dot{E}_\mathrm{free} = \frac{5 \epsilon \dot{M} \left(j-j_\mathrm{PNS}\right)^2}{4 R^2}.
\end{equation}
We explore three different values for the efficiency parameter, $\epsilon = 0.1, 0.5, 1.0$.
This allows us to survey a broader range
of plausible jet energy fluxes that may
occur in more realistic MHD simulations,
e.g., due to variations in initial magnetic field strengths and geometries.

If the baryonic mass $M_\mathrm{core}$ of the excised compact remnant increases beyond $2.5 \,\Msun$, we assume BH formation occurs. In this case, the energy input into the jet is terminated.

The inner boundary conditions for the zones
with an angle of $\theta_\mathrm{jet} = 0.1\,\mathrm{rad}$ of the grid axis are chosen such as to reproduce the desired
total energy flux $\dot{E}_\mathrm{jet}$ for bipolar jets in both hemispheres.
To obtain the correct relativistic energy flux, we require
\begin{equation}\label{eq:jet_energy}
	\frac{\dot{E}_\mathrm{jet}}{2\,\mathrm{d}\Omega_\mathrm{jet}r^2 } = \alpha \phi^6\rho\, (h W^2 -W)\, \hat{v}^1 ,
\end{equation}
where $\mathrm{d}\Omega_\mathrm{jet} = 2\pi|1-\cos\theta_\mathrm{jet}|$, $\alpha$ is the lapse function, $\phi$ is the conformal factor in
the xCFC metric, $\rho$ is the density, $h$ is the relativistic specific enthalpy, and $W$ is the Lorentz factor. Furthermore, $\hat{v}^1$ is defined as $\hat{v}^1 = v^1 - \beta^1/\alpha$, where $\beta$ is the shift vector and $v^1$ is the radial component of the 3-velocity in the Eulerian frame. See \citet{dimmelmeier_2002,dimmelmeier_2005,mueller_2010} for an in-depth presentation of the relativistic equations of hydrodynamics implemented in the \textsc{CoCoNuT} code. In this way, the jet energy is tied to the mass and angular momentum of the accreted material, in the form of the available free energy in differential rotation. To implement the jets, we rearrange Equation~\ref{eq:jet_energy} for $\rho$, 
\begin{equation}\label{eq:jet_rho}
	\rho = \frac{\dot{E}_\mathrm{jet}}{2\, \mathrm{d}\Omega_\mathrm{jet}\alpha \phi^6 (h W^2 -W)\, \hat{v}^1 r^2  } .
\end{equation}
and set $v$, $h$ to match the properties of simulated MHD jets, as described in Section ~\ref{ssec:calibration_method}. The composition of
the ejected jet material also need to be specified.
We set $Y_\mathrm{e} = 0.3$ for the electron fraction in the jet, though detailed nucleosynthetic calculations are performed in post-processing where $Y_\mathrm{e}$ can be varied as a free parameter, as is described in Section \ref{ssec:pp_nucleo}.

\subsection{Calibrating with MHD results}\label{ssec:calibration_method}
Using Equation (\ref{eq:jet_rho}), we can set the density at the base of our jets according to the desired jet energy flux, if the specific enthalpy, pressure and velocity of the jet material is known. We calibrate the boundary enthalpy of our jets using the results of the \texttt{35OC-RO} model calculated by \citet{obergaulinger_2017},
which is an axisymmetric MHD simulation of a magneto-rotational explosion for progenitor model
35OC. 
For calibration of our model we make use of hydrodynamic results from the \texttt{35OC-RO} model, including the temporal evolution of the radial velocity $v_\mathrm{ref}$, pressure $P_\mathrm{ref}$, and density
$\rho_\mathrm{ref}$
inside the jet at a reference radius of $r_\mathrm{ref}=\mathord{\sim}\,  1,\!000\, \mathrm{km}$, from which we can derive the inner boundary values $h_\mathrm{i}$, $P_\mathrm{i}$ and $\rho_\mathrm{i}$ at  $r_\mathrm{i} = 200\,\mathrm{km}$.

First, assuming the jets consist
of an ideal radiation-dominated gas
with adiabatic index $\gamma = 4 / 3$
and expand adiabatically, we have
\begin{equation}
	\frac{P_\mathrm{i}}{\rho_\mathrm{i}^\gamma} = \frac{P_\mathrm{ref}}{\rho_\mathrm{ref}^\gamma} = P_\mathrm{ref}\left(\frac{\rho_\mathrm{i}}{\rho_\mathrm{ref}} \right)^\gamma \label{eq:pi_po_0}.
\end{equation}
Now, assuming that the jets have a constant opening angle $\mathrm{d}\Omega_\mathrm{jet} = \mathrm{d}\Omega_\mathrm{i} = \mathrm{d}\Omega_\mathrm{ref}$ and propagate with velocity $v_\mathrm{jet} = v_\mathrm{i} = v_\mathrm{ref}$ (i.e., little decelaration of the jet), we can equate the mass fluxes at $r_\mathrm{i}$ and $r_\mathrm{ref}$,
\begin{equation}
	\rho_\mathrm{i}\,v_\mathrm{i}\, \mathrm{d}\Omega_\mathrm{i}\,r_\mathrm{i}^2 = \rho_\mathrm{ref} \,v_\mathrm{ref}\, \mathrm{d}\Omega_\mathrm{ref}\,r_\mathrm{ref}^2,
\end{equation}
to obtain
\begin{equation}
\frac{\rho_\mathrm{i}}{\rho_\mathrm{ref}}  = \left(\frac{r_\mathrm{ref}}{r_\mathrm{i}}\right)^2  \label{eq:rhoi_rhoo}.
\end{equation}
Substituting this result into Equation (\ref{eq:pi_po_0}), we find the pressure $P_\mathrm{i}$ at the inner boundary,
\begin{align}
	P_\mathrm{i} &=  P_\mathrm{ref} \left(\frac{r_\mathrm{ref}}{r_\mathrm{i}}\right)^{2\gamma} \label{eq:pi_po},
\end{align}
and by combining Equations~(\ref{eq:rhoi_rhoo}),
(\ref{eq:pi_po}), and the relativistic enthalpy for a radiation-dominated gas, we finally obtain $h_\mathrm{i}$ in terms of known values,
\begin{equation}
	h_\mathrm{i} =  \frac{4P_\mathrm{i}}{\rho_\mathrm{i}} + 1  		= \frac{4  P_\mathrm{ref} \left(r_\mathrm{ref}/r_\mathrm{i}\right)^{2\gamma}  }{\rho_\mathrm{ref} \left(r_\mathrm{ref}/r_\mathrm{i}\right)^2 } + 1 = (h_\mathrm{ref}-1)  \left(\frac{r_\mathrm{ref}}{r_\mathrm{i}}\right)^{2(\gamma-1)}   + 1,
\end{equation}
using geometrised units for all variables.
Once $h_\mathrm{i}$ and $v_\mathrm{i}$
are specified, we can determine the boundary value $\rho_\mathrm{i}$ from the
jet power using Equation~(\ref{eq:jet_rho}),
\begin{align}\label{eq:jet_rhoi}
	\rho_\mathrm{i} &= \frac{\dot{E}_\mathrm{jet}}{\mathrm{d}\Omega_\mathrm{jet}\alpha_\mathrm{i} \phi_\mathrm{i}^6 (h_\mathrm{i} W_\mathrm{i}^2 -W_\mathrm{i}) \hat{v_\mathrm{i}}^1 r_\mathrm{i}^2  } .
\end{align}
Any other boundary values for primitive variables can be determined from $v_\mathrm{i}$, $h_\mathrm{i}$, and $\rho_\mathrm{i}$.

From the results of \citet{obergaulinger_2017} (see Figure \ref{fig:mhd_comparison}), we have adopted values of $\rho_\mathrm{ref} = 1\times10^6\,\mathrm{g\,cm}^{-3}$, $P_\mathrm{ref} = 4\times10^{24}\,\mathrm{erg\,cm}^{-3}$.

We encounter two situations, i.e., very high $\dot{M}$, or very low $\dot{E}_\mathrm{jet}$, where we must modify our prescription in order to maintain the jet outflows. 
To maintain collimation during periods when the accretion rate is exceedingly high during the first $\sim 10$~milliseconds, we set $v_\mathrm{i} = \mathrm{min}(0.5\,c,0.15\,c\cdot(1+\dot{M}/(\Msun\,\mathrm{s}^{-1})))$.
Additionally, for very low $\dot{E}$, we adjust $v_\mathrm{i}$ to enforce $\rho_\mathrm{i} \gtrsim 1.4\times10^7\, \mathrm{g}\, \mathrm{cm}^{-3}$. If $\rho_\mathrm{i}$ falls below this limit, we incrementally decrease $v_\mathrm{i}$, until $v_\mathrm{i}$ approaches zero, at which point the jets are turned off.

\subsection{Post-processing nucleosynthesis}\label{ssec:pp_nucleo}
As described
in Section \ref{sec:coconut}, we follow the evolution of the nuclear composition and subsequent nuclear energy generation using the simple 19-species network
from \citet{weaver_1978} during our hydrodynamics simulations, which is adequate to capture the feedback of nuclear burning, dissociation, and recombination on the energetics of the explosion. To extract detailed nucleosynthetic yields, we implement Lagrangian tracer particles in a post-processing step and calculate the final nuclear composition of our ejecta particles using the \textit{SkyNet} reaction network software library, which contains 7,843 nuclides and 140,000 reactions \citep{lippuner_2017}.

To extract the thermodynamic trajectories of tracer particles from the hydro results, we set a resolution of one tracer particle per ejecta grid cell, where an ejecta grid cell is defined as having positive velocity and positive net total energy \citep[see][Eq. 20]{muller_2017} at the end of each simulation. We then interpolate the properties of the tracer particles from the hydro grid, following their evolution backwards through the simulation time (for more details on this approach see \citealp{wanajo_2018}).
To calculate the detailed nucleosynthesis for each tracer particle, we pass the temperature and density trajectories into \textit{SkyNet}, which is configured to implement reaction rates from the REACLIB database \citep{cyburt_2010}. We assume nuclear statistical equilibrium (NSE) for $T\geq7\,\mathrm{GK}$, and the detailed balance is used to calculate the inverse rates. We extrapolate the density and temperature of the trajectories out to $10^9\,\mathrm{sec}$ assuming homologous expansion, with $\rho \propto r^{-3}$ and $T \propto r^{-1}$.
The initial composition of the tracer particles that originate inside of magneto-rotational jets is not well constrained. Earlier studies that attempt to self-consistently determine the electron fraction of jet material typically find neutron-rich jets with $0.1 \lesssim Y_\mathrm{e} \lesssim 0.3$ \citep{winteler_2012,nishimura_2017,mosta_2018}. The more recent study of \citet{reichert_20}, however, includes detailed neutrino transport and finds that proton-rich conditions can result, with $Y_\mathrm{e} > 0.5$, in addition to the neutron-rich material.  
For tracer particles that originate inside the jet, we explore two different values for the electron fraction,
$Y_\mathrm{e} = 0.3$ and $Y_\mathrm{e}= 0.5$, to represent both a neutron-rich and charge neutral case. Otherwise, for the tracer particles that represent an element of the shocked stellar mantle, we initiate the reaction network with the initial composition interpolated from the hydro output.


\section{Results}\label{sec:results}
We explore three different values of efficiency for converting free energy in differential rotation to jet energy ($\epsilon = 0.1, 0.5, 1.0$), for each of our three progenitor models. 
The model labels denote the progenitor model combined with the energy conversion efficiency parameter, e.g., Model \modelzah is an explosion of the progenitor Model \modelza with a conversion efficiency of $\epsilon = 1.0$.

For each model, we obtain three sets of nucleosynthetic results. One set is taken directly from the hydrodynamic output where the nuclear burning has been approximated with a 19-species network at
high $Y_\mathrm{e}$ and by freeze-out
from NSE at low $Y_\mathrm{e}$, with the electron fraction of the jet set to $Y_\mathrm{e} = 0.3$. The additional two sets are calculated in post-processing, using the thermodynamic histories of Lagrangian tracer particles extracted from the grid, and a detailed reaction network for two values of $Y_\mathrm{e}$ in the jet, i.e., jet $Y_\mathrm{e} = 0.3$
or $Y_\mathrm{e}=0.5$.

\begin{figure}
	\includegraphics[width=\columnwidth]{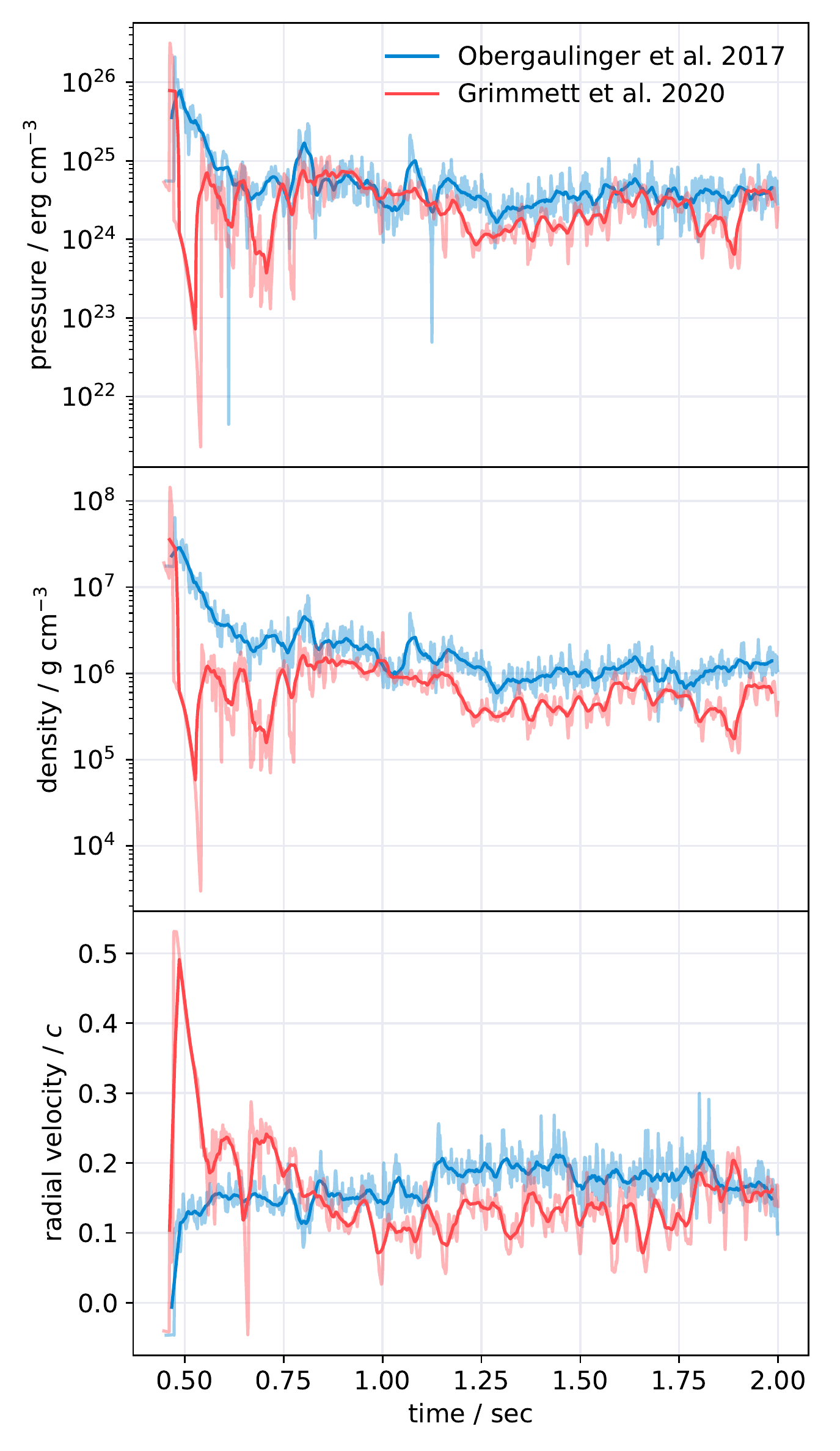}
	\caption{A comparison between our Model \modelocl and the results of \citet{obergaulinger_2017}, for the evolution of pressure, density and radial velocity for a polar zone at $r = 1,\!000\,\mathrm{km}$, top to bottom panels, respectively.  Moving averages are shown by the \textsl{heavy lines} for clarity, the raw data is shown by the \textsl{semi-transparent lines}.}
	\label{fig:mhd_comparison}
\end{figure}

\begin{table}
\caption{The number of tracer particles used for each model.}
\label{table:tracers}
\begin{tabular}{ |c|r|r|r| }
 \hline
 Model
 & Jet & Envelope & Total \\
 \hline
\modelzal & 1,921 & 2,189  & 4,110 \\
\modelzam & 13,777 & 3,497 & 17,274\\
\modelzah & 11,554 & 5,159 & 16,713\\
\modelocl & 1,822 & 1,405 & 3,227\\
\modelocm & 1,650 & 2,287 & 3,937\\
\modeloch & 12,703 & 1,532 & 14,235\\
\modelzbh & 1,352 & 292 & 1,644\\ 
 \hline
\end{tabular}\\
\end{table}

\subsection{Comparison to MHD results}
In Figure \ref{fig:mhd_comparison}, we compare the temporal evolution of $\rho$, $v_\mathrm{r}$, and $P$ for a polar zone (i.e., in the path of the jet) at $r \sim 1,\!000\,\mathrm{km}$ from our Model \modelocl, to the evolution of these properties in the same location in the 2D MHD model \texttt{35OC-RO} of \citet{obergaulinger_2017}.
Overall, we find reasonably good agreement between the two sets of results once the jet outflow has become quasi-stationary. The largest discrepancies occur during the first $\sim100\,\mathrm{ms}$ after the jets are initiated, when the density and pressure in the jets of our Model \modelocl are 2-3 orders of magnitude lower than that of the MHD model, and the radial velocity is a factor of 4-5 larger. During this time our jets are not yet well collimated and expand laterally as they push the dense accretion flow aside and entrain considerable amounts of material into the ejecta, resulting in lower values of pressure and density
compared to the MHD simulation.
After $\mathord{\sim}100\,\mathrm{ms}$ of evolution the jets propagate into less dense material, and the narrow collimation angle is maintained.
During this stage the pressure and density increase toward the intended values, within a factor of $\mathord{\sim}2$ . The jet outflows in the MHD model of \citet{obergaulinger_2017} are collimated by magnetic hoop stresses, and as such the density and pressure in these jets remain a factor of $\mathord{\sim}2$ larger than in our Model \modelocl.

The final outcome found for each model in terms of remnant and explosion energy is also in good agreement. \citet{obergaulinger_2017} find that after $\sim2\,\mathrm{seconds}$ of evolution, their Model \texttt{35OC-RO} has a central remnant with baryonic mass in excess of the assumed BH formation limit ($2.5\,\Msun$), and an explosion energy of $1\times10^{51}\,\mathrm{erg}$. In a similar time frame of evolution for our Model \modelocl, we find that the central remnant mass exceeds the BH limit, and a diagnostic explosion energy\footnote{Here we calculate the diagnostic explosion energy as defined by Eq.~20 of \citet{muller_2017}} of $9\times10^{50}\,\mathrm{erg}$.
The fact that we are able to achieve a reasonable match to Model \texttt{35OC-RO} using a jet model that only converts 10\% of the available free energy in differential rotation to jet energy is consistent with the finding that the explosion of Model \texttt{35OC-RO} is only weakly driven by jets, and represents a transition between the class of neutrino-driven explosions and those driven primarily by MHD jets \citep{obergaulinger_2017,obergaulinger_2020}.

\begin{figure*}
	\includegraphics[width=\linewidth]{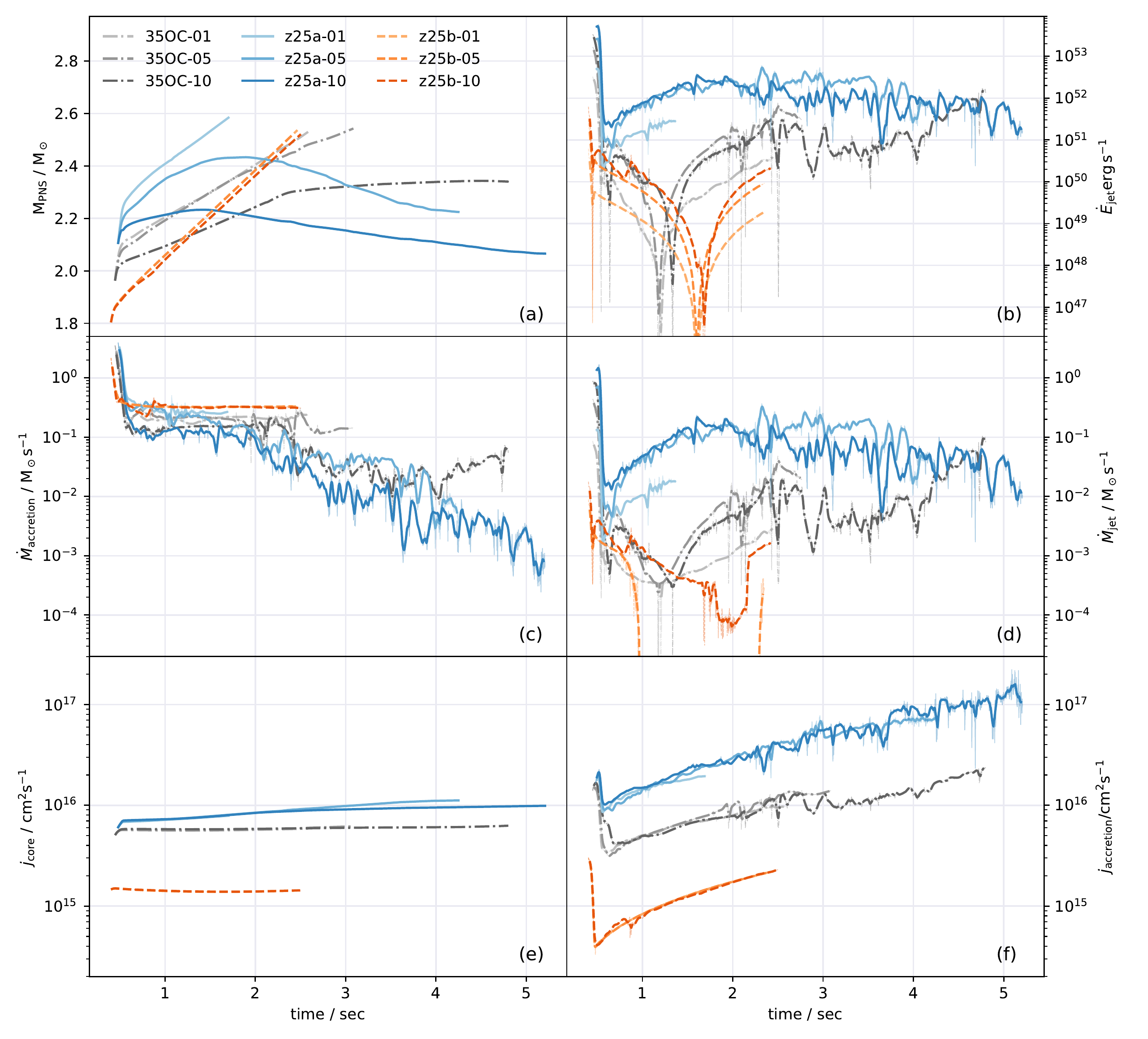}
	\caption{The time evolution of the compact remnant and jet properties for each model, i.e. the flows through the inner boundary. Moving averages are shown for clarity, the raw data is plotted as thin transparent lines. \textit{Panel (a)}: The PNS mass i.e., the mass inside the inner boundary. \textit{Panel (b)}: The jet power, i.e., rate of energy deposition at the polar region of the inner boundary. \textit{Panel (c)}: Rate of mass accretion through the inner boundary. \textit{Panel (d)}: Mass injection rate of the jet. \textit{Panel (e)}: Specific angular momentum of the PNS. \textit{Panel (f)}: Specific angular momentum of the accreting material.}
	\label{fig:global_props}
\end{figure*}

\subsection{Energetics and dynamics of the outflows}
In Figure~\ref{fig:global_props}, we show the evolution of the compact remnant mass, $M_\mathrm{PNS}$, jet power, $\dot{E}_\mathrm{jet}$, mass accretion rate, $\dot{M}_\mathrm{accretion}$, mass ejection rate, $\dot{M}_\mathrm{jet}$ (measured at the inner boundary), specific angular momentum of the core, $j_\mathrm{core}$, and specific angular momentum of the accreting material, $j_\mathrm{accretion}$, for each model.

We find that the maximum jet power (i.e., the maximum power that can be extracted from differential rotation) in our models ranges from $\sim10^{52}\,\ergs$ for the fastest rotating Model \modelza, to $\sim10^{50}\,\ergs$ for the slowest rotating Model \modelzb. This range of values is consistent with the powers found for magneto-rotationally-driven jets that form in models with similar rotation rates \citep[e.g.,][]{akiyama_2003,burrows_2007}. 
The injected jet power depends on the combination of the mass accretion rate, and the difference between the specific angular momentum of the core and the specific angular momentum of the accreting material (see Equation~\ref{eq:e_jet}). This dependence is reflected in the evolution of the jets, where the instantaneous jet power varies as a function of the rate of mass and angular momentum accretion. 
In each model, there is an initial brief peak in jet power as the material near the outer edge of the rapidly rotating core is quickly accreted. Within $\mathord{\sim}50\,\mathrm{ms}$ the jets begin to drive away some of the inner material and slow the accretion rate.
In each of the \modelzb and \modeloc models, the jets are briefly interrupted when $j_\mathrm{accretion} \simeq j_\mathrm{core}$. This can seen as a sharp dip in jet power at $t \sim 1\,\mathrm{s}$ for the Models \modeloc, and $t \sim 1.6\,\mathrm{s}$ for Models \modelzb.  The most energetic models, \modelzah and \modelzam, have driven away much of the inner material after $\sim 2 \texttt{-} 3\,\mathrm{s}$, resulting in a low mass accretion rate and in turn, a decreasing jet power. 

Progenitor \modelzb  contains only marginally sufficient rotational energy to sustain jet outflows. The free rotational energy must be converted to jet energy with 100\% efficiency
as in Model \modelzbh to sustain even weak jets ($\sim10^{50}\,\ergs$) during the $\mathord{\sim}2\,\mathrm{s}$ window before BH collapse. These weak jets have almost negligible effect on the mass accretion rate, and there is little difference in the time to BH collapse between Model \modelzbh and Models \modelzbm/\modelzbl. Models \modelzbm and \modelzbl have a jet power $< 10^{50}\,\ergs$, which is inadequate to overcome the momentum of the infalling material, and so the outflow is quenched entirely. As such, the explosions of the \modelzb progenitor are not compatible with the energetic SN Ic-BL signature, and may belong to another subclass of explosions arising from massive and rotating models. We will include the results for its explosion properties and $^{56}\mathrm{Ni}$ ejecta for reference, but will not analyse these models in detail.

We terminate each model that reaches $M_\mathrm{core} \geq 2.5\,\Msun$, where $M_\mathrm{core}$ is the baryonic mass of the core. It is interesting to note the correlation between the initial value of $M_\mathrm{PNS}$ after collapse and the longevity/power of the jets. We see that the progenitor with the smallest initial value of $M_\mathrm{PNS}$, Model \modelzb, inevitably forms a central BH due to the weakly powered jets, whereas the progenitor with the largest initial value of $M_\mathrm{PNS}$, Model \modelza, is able to avoid BH collapse as long as the efficiency for converting energy in differential rotation to jet energy is $\geq 50\%$.
This is due to the positive correlation between progenitor core mass (and hence PNS mass) and rotation rate \citep{heger_2005}. The most massive cores are formed in more rapidly rotating models, which have more free energy in differential rotation within the inner layers. These models propagate highly energetic jets to power an explosion to avoid BH collapse, despite possessing large initial proto-neutron star masses. 

Naturally, $\dot{M}_\mathrm{jet}$ closely follows $\dot{E}_\mathrm{jet}$.  We see that the drop in jet power for the brief period when $j_\mathrm{acc} \simeq j_\mathrm{core}$ in Models \modeloc and \modelzb is less apparent in $\dot{M}_\mathrm{jet}$, though the material injected during this time is unlikely to possess sufficient momentum to be ejected completely. In the most energetic Models \modelzah, \modelzam, and to some extent Model \modeloch, the large $\dot{M}_\mathrm{jet}$ is comparable to the accretion rate, which itself is reduced due to the powerful nature of the jets, and as a result the core mass is held constant or even decreases. This is similar to the evolution observed in the most energetic models of \citet{obergaulinger_2017,obergaulinger_2020}.

\subsection{$^{56}$Ni production}
Our simulations allow us to address the production
of radioactive $^{56}\mathrm{Ni}$ as a key metric for
connecting to observations of Ic-BL supernovae.
From the results of the post-processing nuclear network, we find that models with jet $Y_\mathrm{e} = 0.5$ typically eject a $1.5 \texttt{-} 2$ times larger mass of $^{56}$Ni than those models with jet $Y_\mathrm{e} = 0.3$. Neutron-rich jet particles are able to synthesise increasingly heavy elements via neutron-capture, and contain negligible amounts of $^{56}$Ni at the end of nuclear burning. The $^{56}$Ni mass ejected from models with jet $Y_\mathrm{e} = 0.3$ is produced entirely within the shocked stellar envelope, which is composed of material with $Y_\mathrm{e} \simeq 0.5$. Models with jet $Y_\mathrm{e} = 0.5$ produce the same mass of $^{56}$Ni in the envelope, plus an additional mass of $^{56}$Ni produced within the jet.

The approximate network that is built into the hydrodynamic solver calculates a $^{56}$Ni ejecta mass that is typically larger than that calculated by the \textit{SkyNet} network for the same jet $Y_\mathrm{e}$. This is because the smaller network cannot accurately follow the neutron-capture reactions and therefore much of the matter that would otherwise be processed to increasingly heavy elements instead remains as $^{56}$Ni. The ejected mass of $^{56}\mathrm{Ni}$ calculated by the approximate network with jet $Y_\mathrm{e} = 0.3$ is on average $\sim30\,\%$ larger than the results of the \textit{SkyNet} network for the same jet $Y_\mathrm{e}$. For a rough estimate of the amount and distribution
of $^{56}$Ni, a small network appears adequate, however.

\begin{figure}
	\includegraphics[width=\columnwidth]{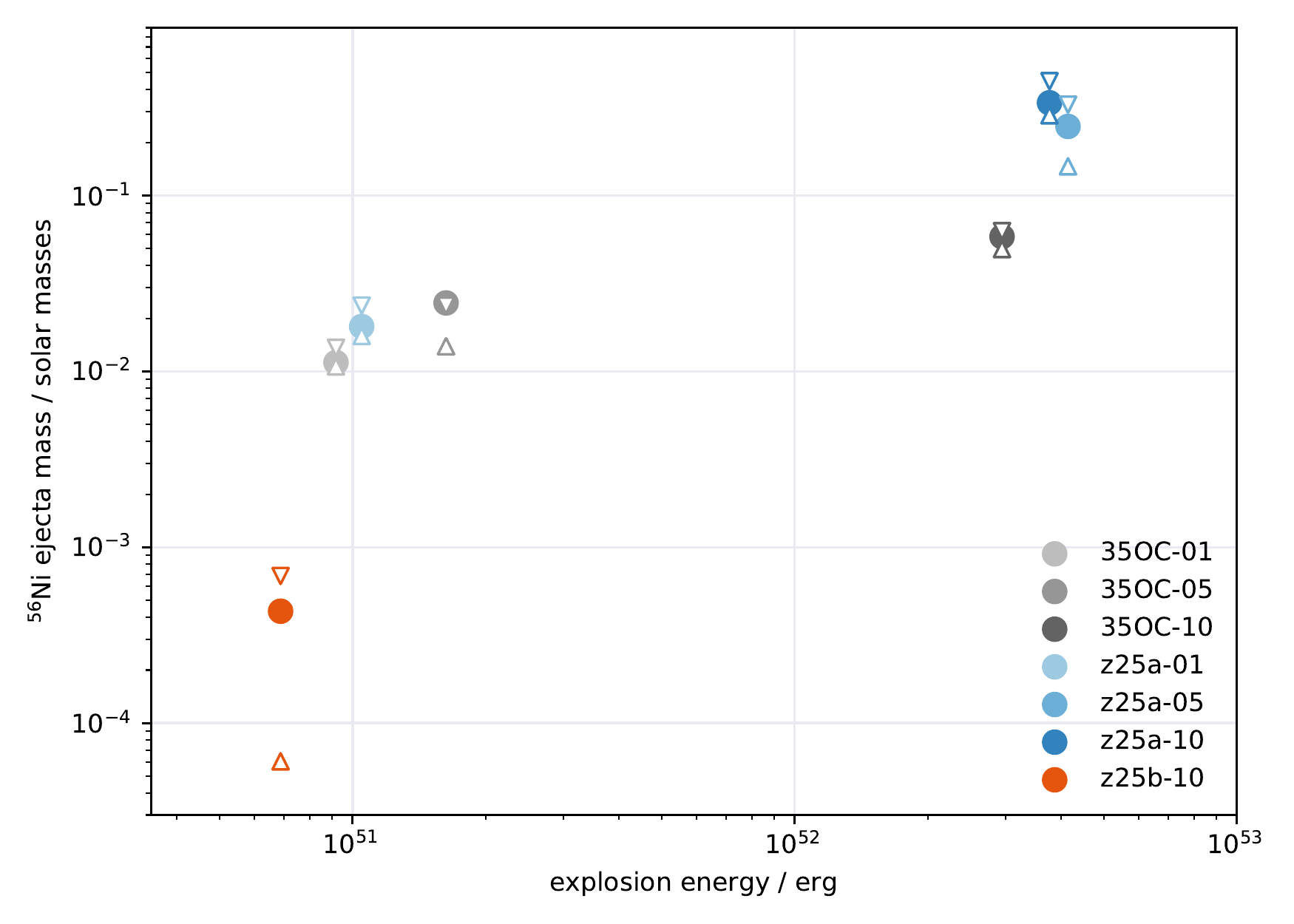}
	\caption{The mass of $^{56}\mathrm{Ni}$ in the ejecta at the end of the simulation time for each model, except Models \modelzbm and \modelzbl, which both fail to explode. The circles are the results of the small 19-species network. The downward and upward pointing triangles are the results taken from the \textit{SkyNet} calculation with jet $Y_\mathrm{e} = 0.5$ and $Y_\mathrm{e} = 0.3$, respectively.}
	\label{fig:ni56_en}
\end{figure}

\begin{figure*}
    \begin{subfigure}{\columnwidth}
        \includegraphics[width=\columnwidth]{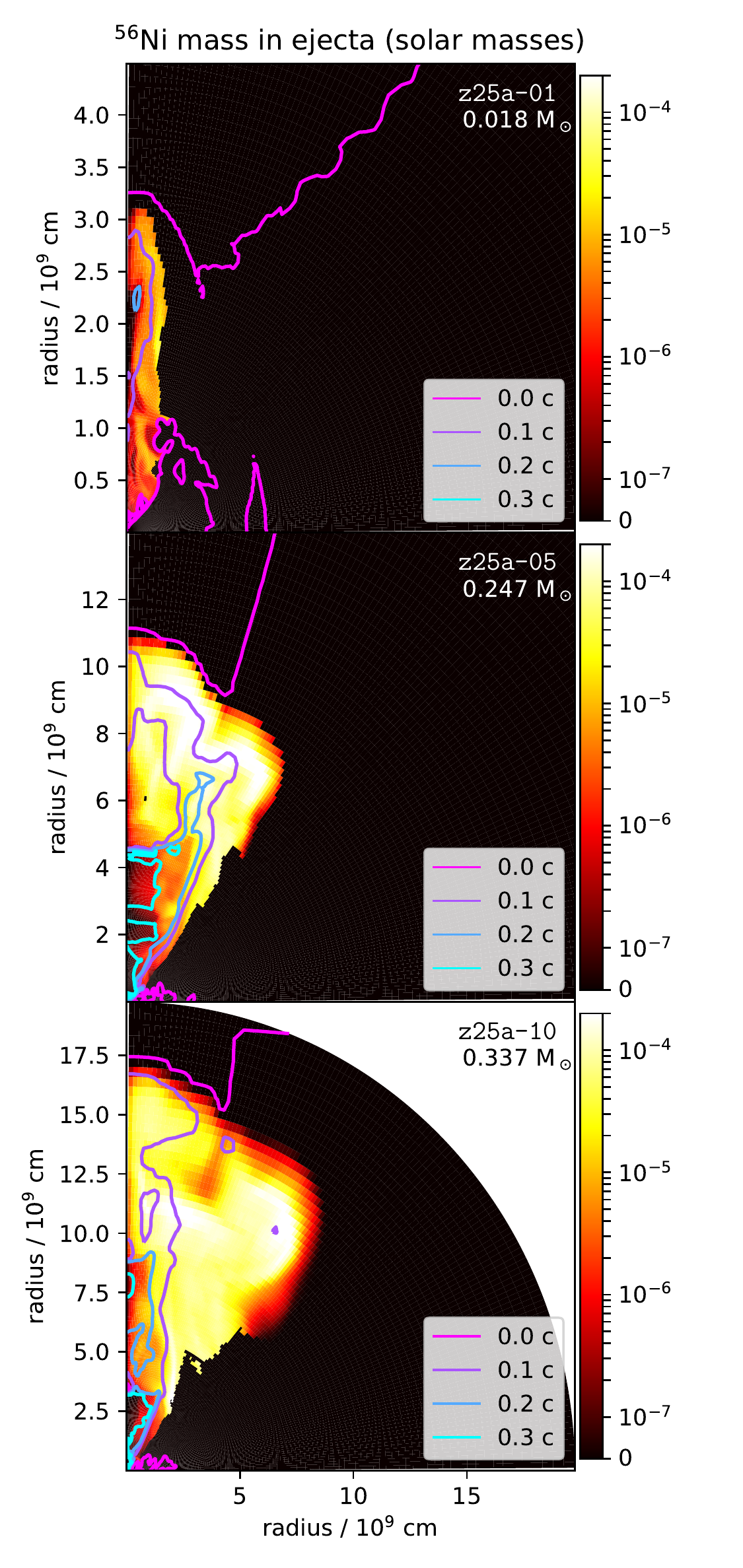}
        \caption{Models \modelza.}
        \label{fig:ni56_vex_z25a}
    \end{subfigure}%
    \begin{subfigure}{\columnwidth}
	    \includegraphics[width=\columnwidth]{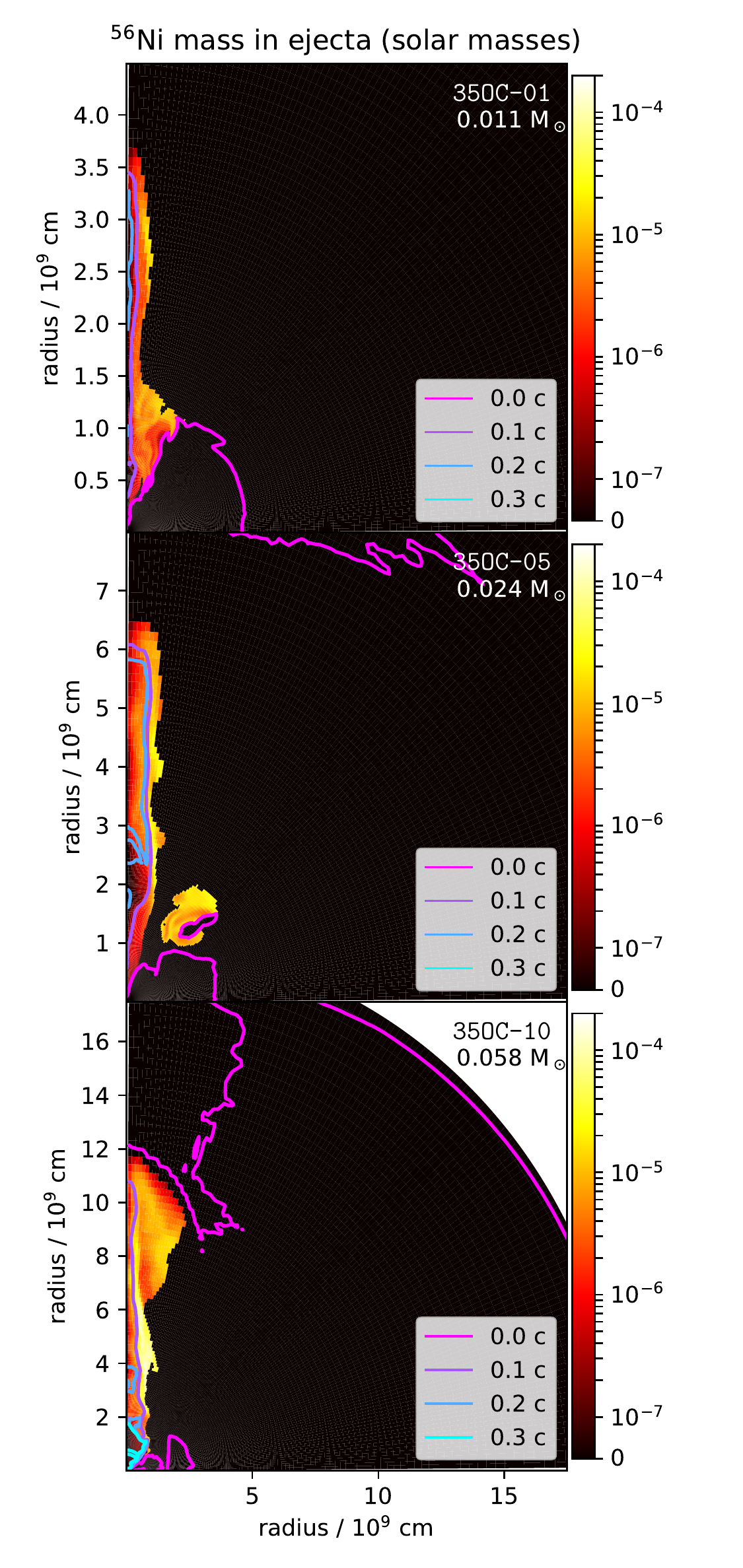}
	    \caption{Models \modeloc.}
	    \label{fig:ni56_vex_35OC}
	\end{subfigure}
	\caption{The distribution of  $^{56}\mathrm{Ni}$ at the end of the simulations, calculated by the 19-species network, in solar masses per grid cell. The contour lines show the velocity of the ejecta in units of $c$.
	The total mass of ${}^{56}\mathrm{Ni}$ is indicated below the model name in the op right of each panel.}
\end{figure*}

We find a maximum $^{56}$Ni ejecta mass of $0.45\,\Msun$ (Model \modelzah with jet $Y_\mathrm{e} = 0.5$), and a (non-zero) minimum of $6\times10^{-5}\,\Msun$ (Model \modelzbh with jet $Y_\mathrm{e} = 0.3$). The ejected mass of $^{56}$Ni is correlated with the explosion energy, as shown in Figure \ref{fig:ni56_en}, which is to be expected as both properties are determined by the power and longevity of the jets. In Figures \ref{fig:ni56_vex_z25a} and \ref{fig:ni56_vex_35OC} we show the location of the $^{56}\mathrm{Ni}$ in the ejecta as calculated by the approximate nuclear network, which shows that the majority of the $^{56}$Ni is produced in the polar direction, where the shock is strongest. The two most energetic models, \modelzam and \modelzah, produce $^{56}$Ni in a relatively broad region due to both lateral expansion of the jet caused by high jet power, and from Kelvin-Helmholtz instabilities which develop along the side of the jet. This lateral expansion may not occur if magnetic fields were present to provide hoop stresses via the twisted toroidal field to collimated the jet, as in the MHD models of, e.g., \citet{wheeler_2002,akiyama_2003,burrows_2007,winteler_2012,mosta_2014,obergaulinger_2020,kuroda_2020}. We also see in Figures \ref{fig:ni56_vex_z25a} and \ref{fig:ni56_vex_35OC} that the $^{56}\mathrm{Ni}$ at the head of the jet is travelling at speeds $\gtrsim0.1c$, and in the most energetic Models \modelzah and \modelzam, the jet has traversed
more than $10\times10^4\,\mathrm{km}$ through the most dense regions of the stellar envelope. 

\begin{figure}
	\includegraphics[width=\columnwidth]{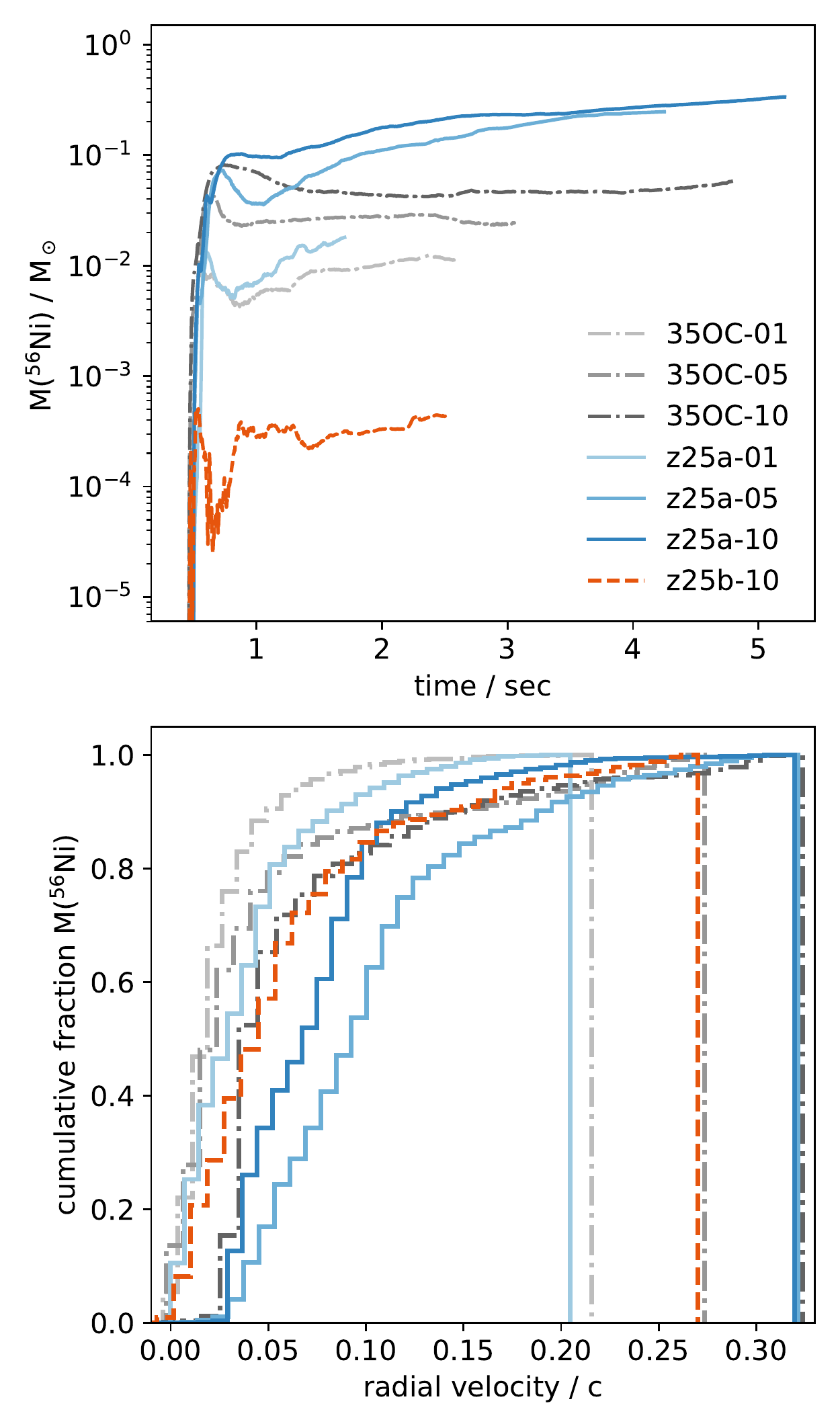}
	\caption{\textit{Upper panel}: the mass of $^{56}\mathrm{Ni}$ in the ejecta of each model as a function of time. \textit{Lower panel}: the cumulative distribution of the total $^{56}\mathrm{Ni}$ ejecta mass as a function of its radial velocity, shown for each model at the end simulation time.}
	\label{fig:ni56_production}
\end{figure}

We see in the top panel of Figure~\ref{fig:ni56_production} that for Models \modeloc and \modelzb, the bulk mass of $^{56}$Ni is created within the first second of jet propagation. The same statement is essentially true for the \modelza series, although these models continue to create additional $^{56}$Ni throughout their entire simulated evolution as a result of the high jet power and mass injection rate. Each model shows a small dip in $^{56}\mathrm{Ni}$ ejecta mass after an initial local maximum in the first $\mathord{\sim}1\,\mathrm{sec}$ of jet propagation. This is because some $^{56}\,\mathrm{Ni}$ is formed close to the inner boundary in the equatorial region by explosive burning in the bow shock, and ends up not being ejected but is accreted shortly thereafter.
Again, these results are taken from the approximate network, which is broadly representative of the results of the \textit{SkyNet} network, as we have shown in Figure \ref{fig:ni56_en}.
In the lower panel of Figure \ref{fig:ni56_production}, we see that each model contains some fraction of the $^{56}\mathrm{Ni}$ ejecta mass travelling at speeds $> 0.1 c$. In particular,
approximately one third of the total $^{56}$Ni mass is travelling with radial speeds $> 0.1 c$ in
the energetic Models \modelzam and \modelzah, and some
$^{56}$Ni velocities are as high as $0.3 c$.

\begin{figure*}
	\includegraphics[width=\linewidth]{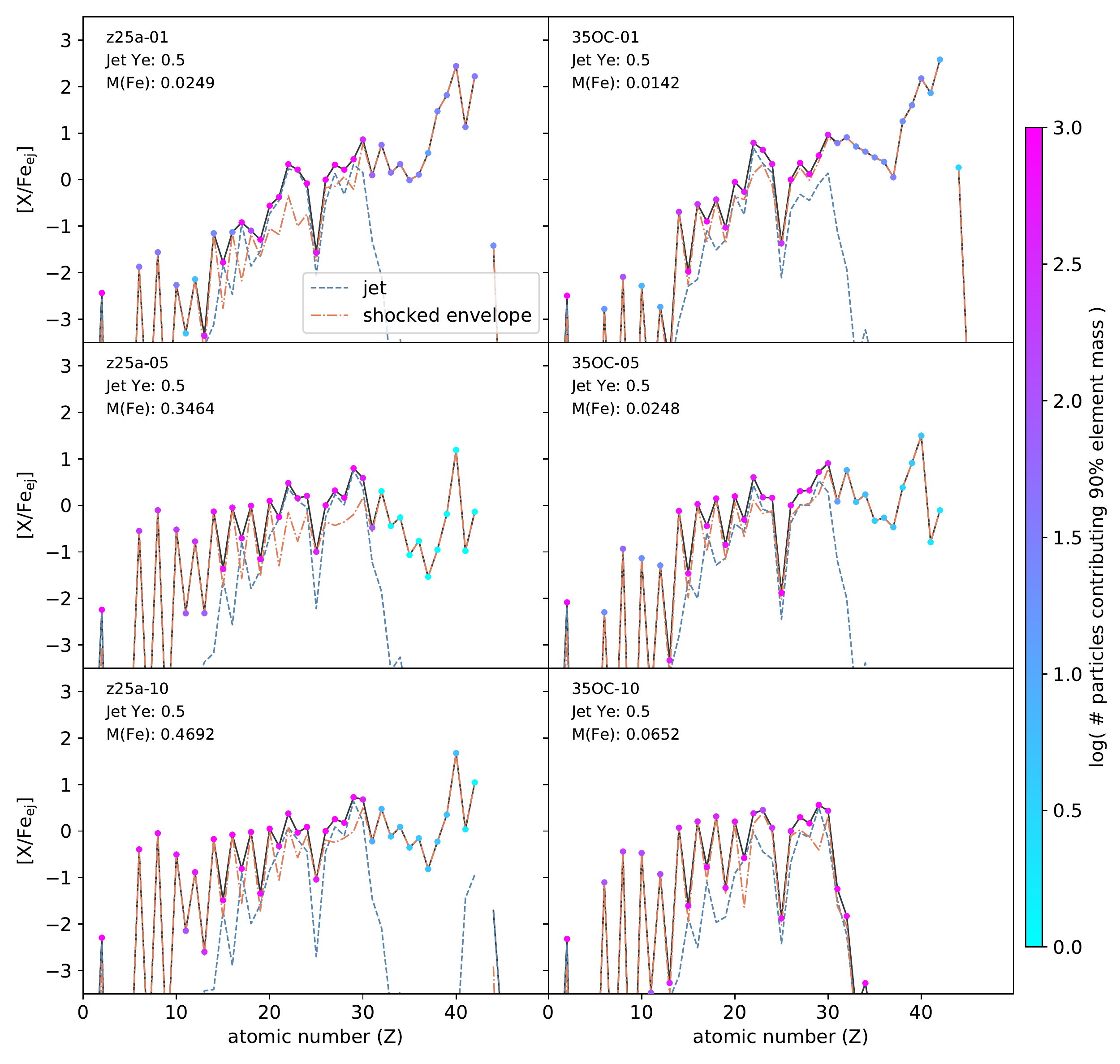}
	\caption{The chemical abundance distributions for each of the \modelza and \modeloc models, as calculated by the \textit{SkyNet} network with jet $Y_\mathrm{e} = 0.5$.  The blue dashed line shows the chemical contribution from tracer particles inside the jet.  The orange dashed line shows the chemical contribution from the tracer particles in material from the shock-heated envelope.  The \emph{solid black line with circle markers} is the combined (jet + envelope) abundance distribution.  Each abundance ratio is normalised by total iron mass in the ejecta (jet + envelope).  The colour of each circle marker indicates the number of tracer particles that contribute $90\,\%$ of the mass yield for each element.}
	\label{fig:abu_grid_ye05}
\end{figure*}

\begin{figure*}
	\includegraphics[width=\linewidth]{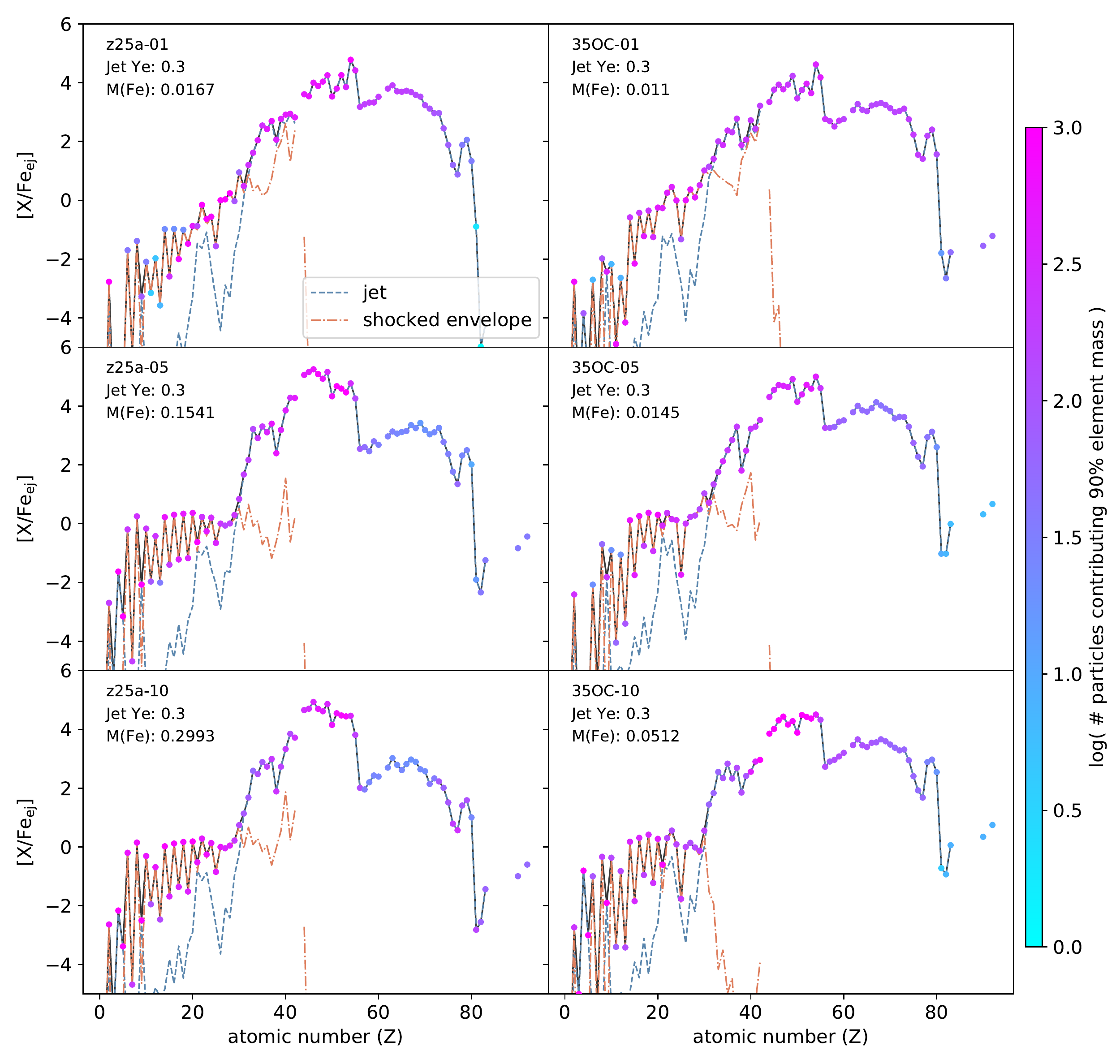}
	\caption{Same as Figure \ref{fig:abu_grid_ye05}, but for jet $Y_\mathrm{e} = 0.3$.}
	\label{fig:abu_grid_ye03}
\end{figure*}

\subsection{Chemical abundance profiles}\label{ssec:nucleo}
In Figure~\ref{fig:abu_grid_ye05} we show the detailed abundance profiles for each of the \modelza and \modeloc models with $Y_\mathrm{e} = 0.5$ in the jet, and in Figure \ref{fig:abu_grid_ye03} the abundance profiles for the same models with $Y_\mathrm{e} = 0.3$ in the jet. The \modelzb Models are excluded from detailed analysis as we already see that they do not eject sufficient $^{56}\mathrm{Ni}$ to be compatible with the SN Ic-BL observational constraints. The chemical yields for each model are computed based on
the mass fractions in the unbound material
only; matter ahead of the shock is not taken into account because it is unclear what fraction
of the envelope will be ejected. How
much of the unshocked envelope material
is ejected primarily affects yields
at $Z\leq 15$. We shall discuss this as
a source of uncertainty later in
Section~\ref{ssec:emp_comparison}.

Within these figures, we also indicate the number of tracer particles contributing significantly to each elemental abundance, and show the separate chemical contributions from the jet and shocked envelope. The abundance ratios making up the separate chemical profiles are each presented as their ratio to the total iron mass in the ejecta, using the standard square bracket notation relative to the solar value, i.e., $[\mathrm{A}/\mathrm{B}] = \log(N_\mathrm{A}/N_\mathrm{B}) - \log(N_\mathrm{A}/N_\mathrm{B})_\odot$.

First, in Figure~\ref{fig:abu_grid_ye05}, we see that the models possessing jets with initial composition $Y_\mathrm{e} = 0.5$ typically produce an abundance pattern that extends beyond the iron-peak elements, containing large abundances of some light neutron-capture elements, e.g., up to Mo ($Z=42$) or Ru ($Z=44$). The exception is Model \modeloch, where chemical abundances drop off rapidly beyond Zn ($Z=30$). From the separated abundance profiles we see that the jet and envelope components each make comparable contributions to elements between P ($Z={15}$) and Zn ($Z=30$), whereas the shocked envelope alone produces the vast majority of elements lighter than P, and those beyond Zn. The elements $Z < 15$ are ejected from the outer layers of the envelope where they were synthesised in part during the nuclear burning lifecycle of the progenitor star and swept up by the jet, and partly by explosive burning in the wake of the bow shock of the jet. The light neutron-capture elements beyond Zn, on the other hand, are produced
deep within the explosion from neutron-rich matter in the silicon shell, which is shocked by and entrained within the jet. In most models,
few (less than ten)
tracer particles contribute to the neutron-capture elemental abundances, or in the case of Model \modeloch, are not present at all. With so few particles representing the abundances of neutron-capture elements in these models, the results beyond $Z=30$ are highly uncertain and will be sensitive to the resolution of the simulation grid and tracer particles. 

In general the yields from the shocked envelope in each model show a strong odd-even effect, whereas the jet component provides a strong enhancement of elements between Ca ($Z=20$) and Z=30,
and exhibits a qualitatively similar shape in all models.
In fact, the abundance patterns up to Zn are remarkably similar for all of the high-energy  models
with $E > 10^{52}\,\mathrm{erg}$
(Models \modeloch, \modelzam, and \modelzah), while the low-energy  models with $E \sim 10^{51}\,\mathrm{erg}$) are also similar to one another, though with somewhat larger variation for elements lighter than Sc ($Z=21$).

The models with lower jet energy efficiency are deficient in the abundances of light and intermediate elements with $Z < 15$. This is because the more energetic models are able to sweep up and eject more of the envelope mass by the end of the simulation time. Presumably, if we were to follow the expansion of the lower energy models (i.e., Models \modelzal and \modelocl) for longer periods of time after BH collapse, the ejected envelope mass and abundances of elements with $Z < 15$ may increase in turn. We explore this assumption in Section~\ref{ssec:emp_comparison}. Due to the large binding energy of
the massive envelope, however, we except that a significant fraction of the un-shocked envelope will undergo fallback for low explosion energies.

Turning our attention to Figure \ref{fig:abu_grid_ye03}, we see that in the case of neutron-rich jets
with $Y_\mathrm{e} = 0.3$, the ejecta
consist almost entirely of heavy neutron-capture elements, and in these models the final abundance profile extends to U
($Z=92$).  The ejected abundances of elements with $Z < 30$ in these models are mainly from the shocked envelope, which is unchanged from the models with jet $Y_\mathrm{e} = 0.5$. The heavy neutron-capture elements are strongly enhanced in these models by the rapid neutron-capture process (\textsl{r}-process) which occurs within the jet, in particular those with $Z=40\texttt{-}55$
(Zr to Cs) are strongly enhanced in each model. The total mass of \textsl{r}-process matter scales with the mass of $^{56}\mathrm{Ni}$ in the ejecta, i.e., between $10^{-3} - 10^{-1}\,\Msun$. 

The final abundance pattern resulting from the \textsl{r}-process is highly sensitive to the value of $Y_\mathrm{e}$ in the material being burnt. Whereas here we are using a fixed value of $Y_\mathrm{e}$ for the jet composition in each model, it is likely true in nature that jets would be composed of material with a range of $Y_\mathrm{e}$ values. The $Y_\mathrm{e}$ value within MHD jets is not well constrained, but recent studies find values ranging between $0.1 \lesssim Y_\mathrm{e} \lesssim 0.3$ \citep{winteler_2012,nishimura_2017,mosta_2018}.  The value of jet $Y_\mathrm{e}$ may vary both within the jet, and throughout the lifetime of the jet.  Whereas it is interesting to find that indeed the \textsl{r}-process does occur within our jet models, and that elements are synthesised out to the third \textsl{r}-process peak ($74 < Z < 83$, though the abundance pattern in each of our models drops of significantly beyond $^{80}$Hg), the precise abundance pattern would vary significantly depending on the value or range of values used for jet $Y_\mathrm{e}$.  In Section \ref{ssec:emp_comparison}, we explore two extra values of jet $Y_\mathrm{e}$ for one model (\modelzam).

\subsection{Comparison to Metal-Poor Stars}\label{ssec:emp_comparison}
\begin{figure*}
  \includegraphics[width=\textwidth]{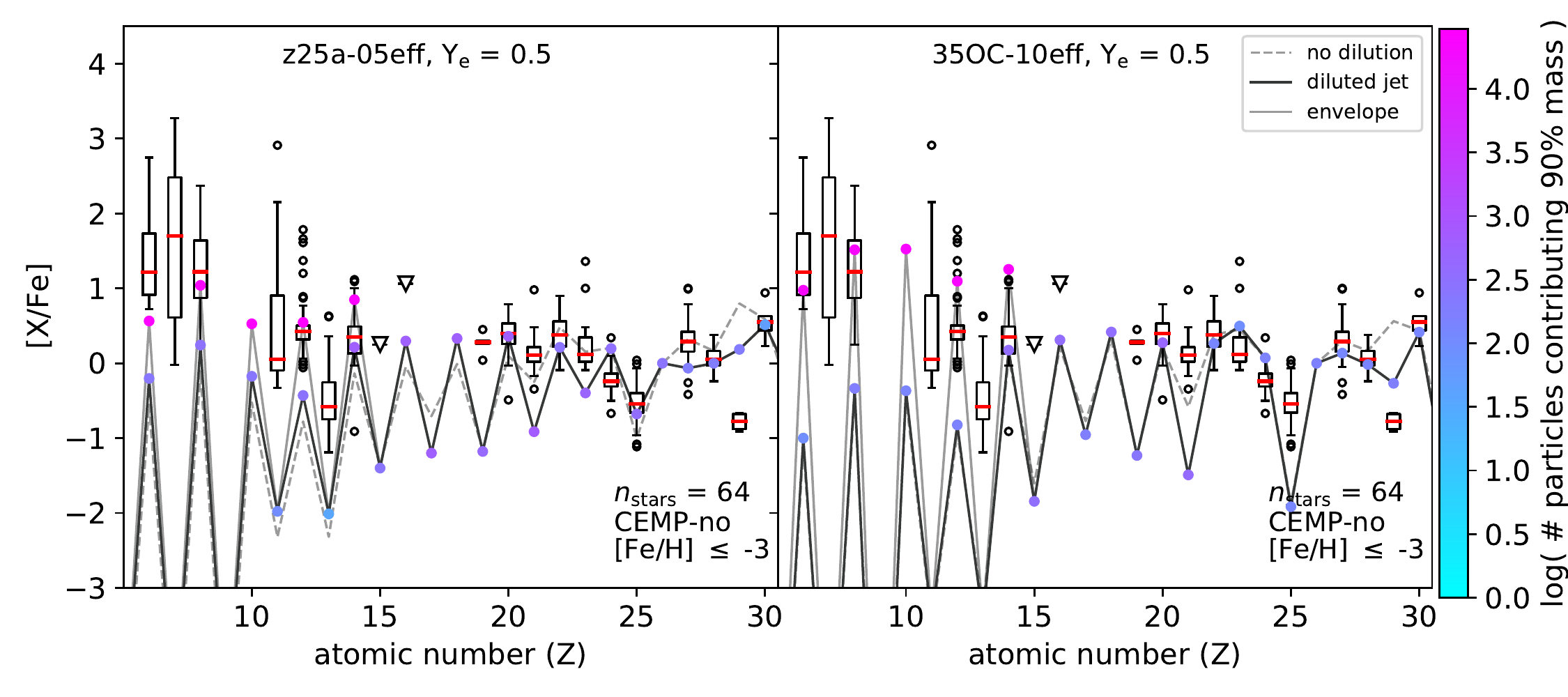}
  \caption{The chemical abundance distributions in the ejecta of Models \modelzam (\textit{left panel}) and \modeloch (\textit{right panel}), each with jet $Y_\mathrm{e} = 0.5$. The \emph{black line with circle markers} shows the chemical abundance ratios in the ejecta, where the jet ejecta mass has been diluted by a factor of 100 relative to the shocked envelope ejecta mass. The \emph{dashed grey line} shows the abundance ratios if the jet component is not diluted. The \emph{light grey line with circle markers} shows the upper limit for the lighter elements if the non-ejecta mass is included. The color of the circle markers indicate the number of tracer particles that contribute 90\% of the total mass of that element. Our model results are compared to the distributions of chemical abundances observed in 64 CEMP-no stars with $\mathrm{[Fe/H]} \leq -3$ (\textit{boxes and whiskers}), with data provided by JINAbase \citep{abohalima_2018}.}
  \label{fig:abu_profile_cempno}
\end{figure*}

\begin{figure*}
  \includegraphics[width=\textwidth]{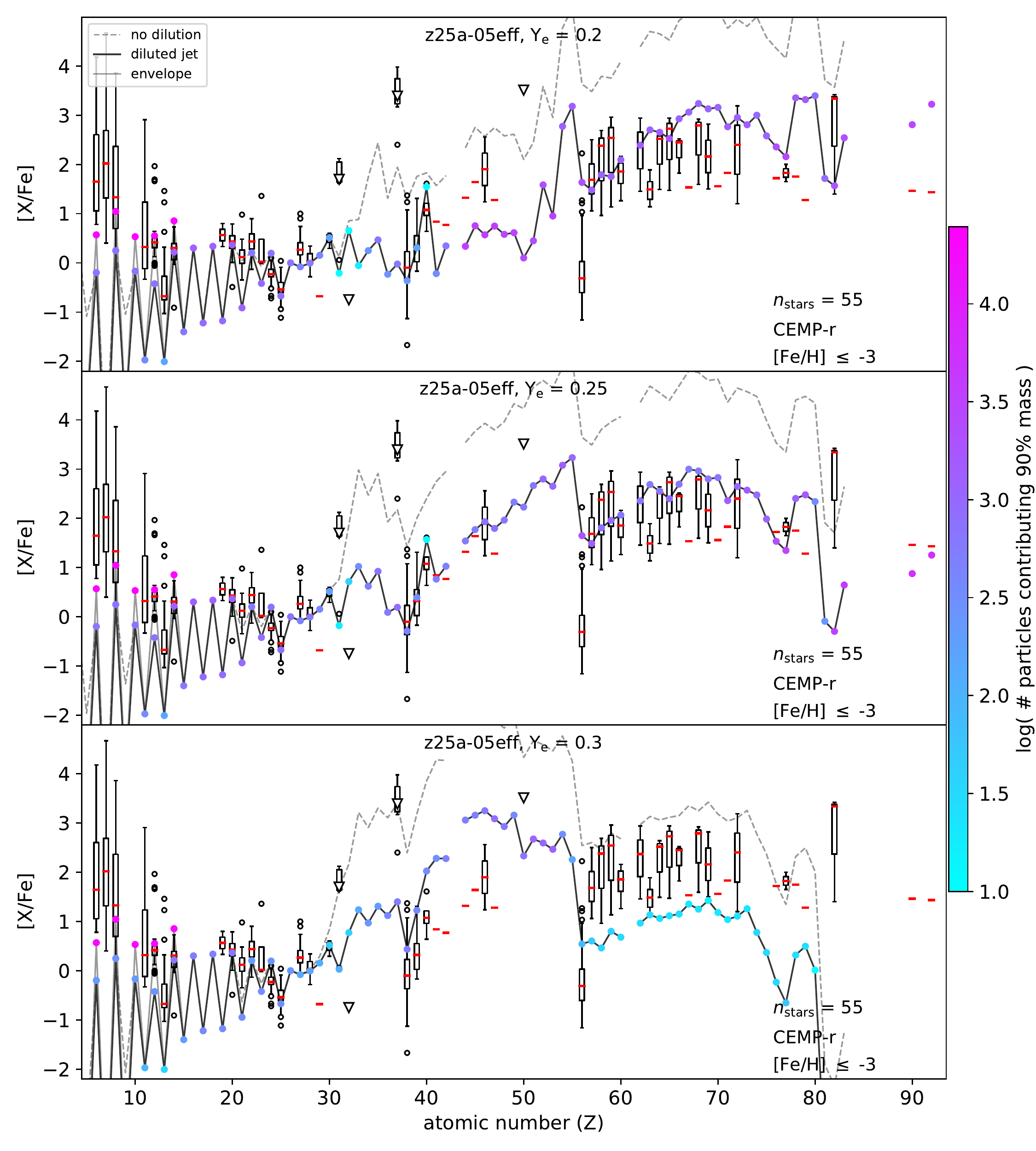}
  \caption{Same as Figure \ref{fig:abu_profile_cempno}, but for the \modelzam model with jet $Y_\mathrm{e} = 0.2$ (\textit{upper panel}), jet $Y_\mathrm{e} = 0.25$ (\textit{middle panel}), and jet $Y_\mathrm{e} = 0.3$ (\textit{lower panel}), compared to 55 CEMP-\textsl{r} stars with $\mathrm{[Fe/H]} \leq -3$.}
  \label{fig:abu_profile_cempr}
\end{figure*}

In Figures~\ref{fig:abu_profile_cempno} and \ref{fig:abu_profile_cempr} we compare a representative selection of our models to the chemical abundances observed in a sample of carbon-enhanced EMP (CEMP) stars.  To provide constraints, we present two sets of yields for our models as limiting cases in which the unshocked material is either completely excluded as in the previous section, or completely included. We have applied a dilution factor of 100 to matter ejected by the jet in order to obtain a reasonable comparison to the observed abundances of \textsl{r}-process elements. We also show the undiluted abundances for reference. The validity of selective and substantial dilution is discussed in Section \ref{sec:discussion}.
We have extracted the stellar observational data from the JINAbase online database of stellar chemical abundances \citep{abohalima_2018}. The CEMP subclass of metal-poor stars bare similar abundance patterns to regular EMP stars, but also have $\mathrm{[C/Fe]} \geq 0.7$. Like all EMP stars, CEMP stars appear to require an enrichment from a high entropy environment in order to explain their iron-peak abundance ratios \citep{nomoto_2013}. The enhanced [C/Fe] observed in these stars is more likely, however, to be provided by a low-energy event that creates only small amounts of iron. Amongst other possibilities, aspherical hypernovae with a strong polar shock and weaker equatorial explosion are speculated to be the source of chemical enrichment for CEMP stars \citep{tominaga_2009,ezzeddine_2019}. Our models with neutron-rich jets are compared to CEMP stars that exhibit an \textsl{r}-process enhancement (i.e., CEMP-\textsl{r} stars, identified by $\mathrm{[Eu/Fe]} > 1$), and models that have jets with $Y_\mathrm{e} = 0.5$ are compared to stars observed without a neutron-capture enhancement (i.e., CEMP-no, identified by $\mathrm{[Ba/Fe]} < 0$).  See \citet{beers_2005,aoki_2007} for further discussion of the subclasses of CEMP stars.
We show the distribution of stellar abundance ratio measured for each element with respect to iron with a box-and-whisker plot in Figures \ref{fig:abu_profile_cempno} and \ref{fig:abu_profile_cempr}.  In the case that there are multiple measurements of an elemental abundance for the same star from different literature sources, we take the average value. We exclude the values of upper limit measurements from the distribution, except when \emph{only} upper limit measurements are available for a particular element, in which case we show the distribution of upper limits.

Figures~\ref{fig:abu_profile_cempno} and \ref{fig:abu_profile_cempr} show that Model \modelzam does not produce sufficiently large [C/Fe] to match the observed distribution in CEMP stars, even if the upper limit value including non-ejected carbon is considered. In the right panel of Figure \ref{fig:abu_profile_cempno}, however, Model \modeloch does have an upper limit to [C/Fe] that falls within the observed distribution. The difference in [C/Fe] between these models is caused by the larger iron production in the more energetic \modelzam model, rather than a difference in carbon production. Model \modelzam ejects $0.35\,\Msun$ of iron and $0.2\,\Msun$ of carbon (upper limit $1.19\,\Msun$), whereas Model \modeloch ejects $0.065\,\Msun$ of iron and $0.011\,\Msun$ of carbon (upper limit $1.0\,\Msun$), resulting in a larger [C/Fe] value in Model \modeloch. 

The abundances of elements 
with $Z=20\texttt{-}30$ (Ca to Zn) are well explained by our models. This can most clearly be seen in Figure \ref{fig:abu_profile_cempno}, as the ejected abundances up to $Z=30$ are identical between the models for different jet $Y_\mathrm{e}$ given the applied degree of jet dilution, nor do the observed abundances in this range change substantially between CEMP-no and CEMP-r stars. The \modelzam model provides a particularly good fit to the observed abundances, and only $\mathrm{Sc}$
($Z=21$) and V ($Z=23$) are underproduced relative to the observed values. The undiluted jet contribution with $Y_\mathrm{e} = 0.5$ would bring these two elements within the observed distribution. There is no justification, however, for only applying a dilution to the neutron-rich jets, and it is clearly necessary to reduce the \textsl{r}-process enrichment in the neutron-rich models. Cu ($Z=29$) is significantly overproduced, though the copper abundance is only measured in four stars, so the statistics are poor. Most notably our models consistently produce high [(Co,Zn)/Fe] values, which is historically a difficult feature to explain in EMP stars.

In Figure~\ref{fig:abu_profile_cempr} we see that the elements beyond $Z=30$ and below the second \textsl{r}-process peak ($48 < Z < 59$) are under-produced by Model \modelzam with jet $Y_\mathrm{e} = 0.2$, and overproduced by the same model with jet $Y_\mathrm{e} = 0.3$. On the other hand, many of the elements beyond the second \textsl{r}-process peak are marginally overproduced by the model with jet $Y_\mathrm{e} = 0.2$, and underproduced by the same model with jet $Y_\mathrm{e} = 0.3$. 
The neutron-capture elements beyond $Z=30$ and up to $Z=92$ are best fit by a model with jet $Y_\mathrm{e} = 0.25$.
Several elements are still poorly fit, most notably Ba ($Z=56$) and Pb ($Z=82)$, although some outlying observed values of Ba are compatible with our model.

\section{Discussion}\label{sec:discussion}
We now evaluate our results in the context of observational constraints provided by SNe Ic-BL and the chemical abundances in metal-poor stars. Some constraints emerge on the parameters of jet-driven models based on the millisecond-magnetar scenario.

\subsection{Broad-lined Ic Supernovae}
The bright peak luminosity and broad spectral lines observed in SNe Ic-BL imply a fast-moving (approaching $\sim 30,\!000\,\mathrm{km\,s}^{-1}$, or $0.1c$) and massive $^{56}\mathrm{Ni}$ ejecta. \citet{taddia_2019} estimate the $^{56}$Ni ejecta mass for 24 SNe Ic-BL and report a distribution between $0.12\,\Msun - 0.80\,\Msun$, with median $0.28\,\Msun$. 

In agreement with \citet{nishimura_2015,chen_2017} we find that a negligible mass of $^{56}\mathrm{Ni}$ is created by neutron-rich ($Y_\mathrm{e} \lesssim 0.3$) jets. Instead, we find that the majority of the $^{56}\mathrm{Ni}$ is synthesised in the shock-heated envelope. If the electron fraction in the jet matter is larger ($Y_\mathrm{e} = 0.5$), however, then the mass of $^{56}\mathrm{Ni}$ created in the jet can be comparable to the mass synthesised in the shocked envelope ejecta. Similar to \citet{maeda_2003,chen_2017}, we find that jet power $\mathord{\sim} 10^{51}\,\mathrm{erg\,s}^{-1}$ is insufficient to produce $M(^{56}\mathrm{Ni}) > 0.1\,\Msun$, and only our models that reach $\dot{E}_\mathrm{jet} \gtrsim 10^{52}\,\mathrm{erg\,s}^{-1}$ produce such large quantities of $^{56}\mathrm{Ni}$.  

Our three most energetic models ($E > 10^{52}\,\mathrm{erg}$), \modelzah, \modelzam, and \modeloch, eject\footnote{The value enclosed in brackets indicates the additional $^{56}\mathrm{Ni}$ mass in the ejecta if the jet has composition $Y_\mathrm{e} = 0.5$} $0.29(+0.16)\,\Msun$, $0.15(+0.18)\,\Msun$, and $0.05(+0.01)\,\Msun$ of $^{56}$Ni, respectively. 
These results overlap with the lower half of the observed distribution reported by \citet{taddia_2019}, but cannot explain the larger $^{56}\mathrm{Ni}$ masses, particularly if the jets are neutron-rich. The $^{56}\mathrm{Ni}$ ejecta masses inferred from observations are calculated under the assumption of spherical symmetry, however, and therefore the observational $^{56}\mathrm{Ni}$ masses may be systematically overestimated if the explosions observed as SNe Ic-BL are indeed intrinsically aspherical. 

A large fraction ($20\%$ -- $40\%$) of the $^{56}\mathrm{Ni}$ mass in energetic Models \modelzah, \modelzam, and \modeloch, is travelling faster than $0.1c$ at the end of our simulation time, after the shock has already traversed a distance of up to $\mathord{\sim}1.5\times10^5\,\mathrm{km}$ through the stellar envelope. 
The velocities are are thus comparable to
or even slightly larger than the outer
ejecta velocities in SNe Ic-BL \citep{modjaz_2016}.
These models, therefore, may produce an observational signature that is similar to those observed in SNe Ic-BL, if the velocity of the ejected $^{56}\mathrm{Ni}$ can be maintained. This is not unlikely given the decreasing density profile in the outer envelope. 

Due to the positive correlation between progenitor core mass and rotation rate \citep[e.g.,][]{heger_2005}, we have found that models with more massive iron cores have more rotational energy available to power jets. As a result, the models that are able to explode energetically and eject the envelope may have a tendency to leave relatively massive neutron star remnants. Some of the initial core mass can also be lost, however, if the jet mass ejection rate exceeds the accretion rate.

Our lower energy ($\mathord{\sim}10^{51}\,\mathrm{erg}$) Models \modelocm, \modelocl, and \modelzah, each eject $M(^{56}\mathrm{Ni}) \lesssim 0.02\,\Msun$. The short-lived jet outflows in these models are unable to appreciably reduce the momentum of infalling material and therefore the accretion rate, and so the mass of the PNS in these models quickly exceeds the BH limit. Hence the $^{56}\mathrm{Ni}$ synthesised in these models is still deep within the envelope ($r = 3\times10^4\,\mathrm{km} - 6\times10^4\,\mathrm{km}$) when the jet energy injection is switched off. So, while the $^{56}\mathrm{Ni}$ mass is insufficient to explain the typical value inferred for SNe Ic-BL, it is less likely that these models would produce a broad-lined spectral signature in any case. This could suggest
that the distribution of magneto-rotational
explosions extends into energetically normal
SNe Ic.

On the other hand, the ongoing accretion of the envelope onto the central BH may set up an environment that allows for a collapsar-type explosion during the later evolution.
Such hybrid hypernova explosions, with weak early magneto-rotational jets and later stage collapsar-type jets have been suggested by \citet{obergaulinger_2017,obergaulinger_2020}. Future work into these types of models will certainly yield interesting results, as the later stage collapsar jets would add an additional mass of $^{56}\mathrm{Ni}$ for powering the luminosity, and other heavy elements to the final chemical abundance pattern. The mass of $^{56}\mathrm{Ni}$ shown to be created by jets in a collapsar environment is similar to the mass ejected by our more energetic models, i.e., $> 0.1\,\Msun$ \citep[e.g.,][]{maeda_2003,barnes_2018}.
Moreover, the collapsar jets would be travelling through a lower density medium, and may be capable of producing a GRB. 

\subsection{Abundances in Extremely Metal-Poor Stars}
The bulk of the matter contributing to the elemental abundances with $Z \leq 30$ is created in the shock-heated envelope, and all of our models provide a relatively good agreement with the intermediate-mass and iron-peak abundance ratios observed in metal-poor stars, regardless of our chosen jet composition. 
In particular, we find that our models consistently produce the enhanced [Zn/Fe] that is required in order to explain the abundances observed in stars with $\mathrm{[Fe/H]} \lesssim -3$.
Zinc is produced abundantly in the shocked-heated envelope, and the [Zn/Fe] in the ejecta therefore is relatively independent of the jet composition. Even if the jet contribution is subtracted, the smallest [Zn/Fe] is produced by Model \modeloch with $\mathrm{[Zn/Fe]} = 0.4$. The largest [Zn/Fe] we find is in Models \modelocl and \modelocm with $\mathrm{[Zn/Fe]} = 1.0$. These results are in good agreement with the range of [Zn/Fe] abundances observed in metal-poor stars.

The models with neutron-rich jets, however, overproduce the \textsl{r}-process elements by a factor of $100$, when compared to the [X/Fe] neutron-capture abundances observed in \textsl{r}-process enhanced metal-poor stars. For each model with neutron-rch jets, the total \textsl{r}-process mass is of the same order of magnitude as the mass of $^{56}\mathrm{Ni}$ ejecta. The total mass of \textsl{r}-process matter that we calculate is similar to that found in the neutron-rich MHD jets of \citet{winteler_2012,nishimura_2015,nishimura_2017,mosta_2018} (i.e., $10^{-3}\,\Msun - 10^{-2}\,\Msun$), though our energetic Models \modelzam and \modelzah eject $\mathord{\sim}10^{-1}\,\Msun$ of \textsl{r}-process matter, in proportion to the larger $^{56}\mathrm{Ni}$ ejecta mass. A total mass of \textsl{r}-process matter that is similar to the mass of $^{56}\mathrm{Ni}$ ejecta results in $\mathrm{[X/Fe]} > 3 - 4$ for the neutron-capture elements. 

One could consider a scenario where the chemical yields from our models are diluted with the yields of nearby CCSNe prior to subsequent star formation in order to bring the [r/Fe] values down to $\mathrm{[r/Fe]}\mathord{\sim}2$. This would would require \textit{at least} an order of magnitude larger Fe mass than is produced by a single one of our HN models (i.e., an increase from $\mathord{\sim}0.1\,\Msun$ to $\mathord{\sim}1.0\,\Msun$). Under the assumption that the typical amount of Fe is produced per CCSNe (i.e., $0.01\,\Msun$), each of our HN models would need to be diluted with approximately 100 regular CCSNe. The frequency of HNe in the early Universe is speculated to be as high as $>$~10 percent, however \citep{podsiadlowski_2004,woosley_2006,arcavi_2010,smith_2011,smidt_2014}. Moreover, if our HNe yields were diluted to such a degree, they would no longer provide the enhanced [Zn/Fe] observed in many metal-poor stars.

This strongly suggests that magnetorotational
explosions cannot involve sustained
neutron-rich jets. As the explosion energy
builds up during the non-relativistic phase
of the jet evolution, the ejection of matter
from the proto-neutron star surface must be 
likely slow enough to allow neutrino processing to
reach values of $Y_\mathrm{e}\approx 0.5$. In future studies, if artificial jets are used and the $Y_\mathrm{e}$ cannot be determined self-consistently, it may be beneficial to consider a distribution of $Y_\mathrm{e}$ values within the jet, perhaps by considering the equilibrium $Y_\mathrm{e}$ value as a function of the neutrino energy and luminosity, along with an appropriate freeze-out radius \citep[see, e.g., Eq. 5 of ][]{muller_2016}.
Theoretically, the problem of producing
excessively high \textsl{r}-process abundances in the
subsequent generations of stars enriched by the hypernova could be avoided by considerably diluting only
the \textsl{r}-process material, e.g., if it mixes
inefficiently with the interstellar medium
and largely escapes the galactic potential.

It is not clear whether the degree of dilution that we have applied to the neutron-rich jet matter in our models is realistic.
Studies into the dispersion of supernova ejecta have indeed shown that 
mixing, turbulence and complex chemistry can lead to differential metal transport, which will ultimately complicate the mapping between Population III chemical yields and the abundance ratios observed on the surfaces on metal-poor stars \citep[e.g.,][]{ritter_2015,ji_2015,chiaki_2019,magg_2020,komiya_2020}.
The highly energetic and aspherical explosions resulting from magneto-rotational models will add further complication to this problem, and further study will be needed to provide constraints on the reasonable amount of absolute and differential dilution.
Jets with relatively high $Y_\mathrm{e}$, however, appear a more natural and
plausible way of avoiding an overproduction
of \textsl{r}-process material than extreme 
assumptions about metal transport in the
interstellar medium.

The large mass of $^{56}\mathrm{Ni}$ (decaying to $^{56}\mathrm{Fe}$) that is ejected from our most energetic models prevents the production of the $\mathrm{[C/Fe]} > 0.7$ that is observed in the majority of the most metal-poor stars \citep{placco_2014,yoon_2018}. 
In general, if a hypernova produces $0.1\,\Msun$ of iron, then all of the $\mathord{\sim}1\,\Msun$ of carbon in the envelope must be ejected in order to have $\mathrm{[C/Fe]}\gtrsim0.7$. If more iron is produced, extra carbon must also be produced in proportion. 
There is no
pathway for substantial carbon production during explosive burning, which indicates that in order to explain the enhanced carbon enrichment with hypernova models alone, no more than $0.1\,\Msun$ of iron can be ejected, and all of the carbon must be ejected from the envelope. If it turns our that $> 0.1\,\Msun$ of iron is consistently produced in the ejecta of hypernovae, then in order to provide an enhanced [C/Fe] value, the progenitor envelope must already by carbon-rich at the end of its nuclear burning lifetime, and ejected along with the larger mass of iron. Otherwise, iron must be diluted relative to carbon prior to subsequent star formation. If none of these scenarios can be fulfilled, then some other source of enrichment is required to explain the enhanced carbon in CEMP stars. It has been suggested, for example, that the origin of enhanced surface carbon abundances may be a result of mass transfer from a rotating binary companion \citep{meynet_2006,maeder_2015}. This scenario is supported by some evidence of a correlation between binarity and absolute carbon abundance in CEMP stars \citep{arentsen_2019}. Alternatively, faint Population III supernovae (perhaps similar to our Model \modelzbh) have been implicated as a potential source of high [C/Fe] enrichment, due to their low iron yield \citep{umeda_2005,tominaga_2007}. Finally, \citet{frebel_2007a,hartwig_2019} have suggested that typical [C/Fe] abundances may be enhanced as a result of inhomogenous mixing of supernova ejecta, and that a shorter cooling time of carbon-rich gas could provide a bias for the formation of carbon-enhanced stars in the metal-poor environment.

\subsection{Evaluation against the combined observational constraints}
The potential of jet-driven hypernova models, under the specific assumptions and parameterisations that we have made, to simultaneously explain the observational
sigantures of SNe Ic-BL and the chemical abundances observed on the surfaces of metal-poor stars can be summarised by the following key points:
\begin{itemize}
    \item Jet-driven hypernovae can consistently produce chemical abundance ratios similar to those observed in metal-poor stars with $\mathrm{[Fe/H]}\leq -3$ for the intermediate-mass and iron-peak elements, and of particular note, $\mathrm{[Zn/Fe]} \geq 0.5$.
    \item Neutron-rich jets overproduce the [X/Fe] abundances of the neutron-capture elements by a factor of 100 compared to those observed in some extremely metal-poor stars. This suggests that the jets can only remain neutron-rich for a short phase, possibly around the onset
    of the magneto-rotational explosion.
    \item Jet-driven hypernovae can produce the large ($>0.1\,\Msun$) mass of $^{56}\mathrm{Ni}$ inferred from the light curves of SNe Ic-BL, and up to $M_\mathrm{ej}(^{56}\mathrm{Ni}) = 0.45\,\Msun$. A high $Y_\mathrm{e}$ in the jet is also preferred because it adds a substantial
    contribution of $^{56}\mathrm{Ni}$ from the jet.
    \item There is, however, a tension between the requirement to produce a large mass of $^{56}\mathrm{Ni}$ needed to explain the SNe Ic-BL light curve, and the capacity of the models to also explain the enhanced carbon abundances ($\mathrm{[C/Fe]}\geq 0.7$) observed in the majority of the most metal-poor stars. The ejected mass of carbon is limited to the amount that is already present in the envelope prior to collapse, and if $M_\mathrm{ej}(\mathrm{Fe}) \gtrsim 0.1\,\Msun$, then $\mathrm{[C/Fe]}\geq 0.7$ requires $M_\mathrm{ej}(\mathrm{C}) \gtrsim 1\,\Msun$.
\end{itemize}

\section{Conclusions}
\label{sec:summary}
We have modelled the collapse and jet-powered explosions of three massive, rotating,  low- or zero-metallicity stars. Motivated by the results of numerical MHD studies, we implement a simple analytic model to relate the jet power to the energy available in the differential rotation of the accreting PNS, and we artificially inject this energy from an inner boundary to launch jets. 
Jet-driven explosions are implicated as the physical counterpart of the SN Ic-BL observational signature, and also as the source of the chemical enrichment observed on the surfaces of metal-poor stars with $\mathrm{[Fe/H]} \leq -3$. 

We find that the energy available to power jets is of the order $10^{50}\,\mathrm{erg\,s}^{-1} - 10^{52}\,\mathrm{erg\,s}^{-1}$, depending on the angular momentum of the progenitor model. Jet power $>10^{51}\,\mathrm{erg\,s}^{-1}$ is sufficient to drive an energetic explosion
with $E > 10^{52}\,\mathrm{erg}$, but jet power $\mathord{\sim}10^{52}\,\mathrm{erg\,s}^{-1}$ is required to produce the large mass
of $>0.1\,\Msun$ of
$^{56}\mathrm{Ni}$  inferred from the observations of SNe Ic-BL. 
Although the models have not been evolved beyond shock breakout, the velocity of the synthesised
$^{56}\mathrm{Ni}$ and its geometric distribution
appear broadly compatible with the findings
from  supernova spectroscopy and spectropolarimetry.
We find that our models can reliably reproduce the chemical abundance ratios observed in metal-poor stars up to [Zn/Fe], with the exception of the enhanced [(C,N,O)/Fe] commonly observed in the most metal-poor stars, which cannot be reproduced due to the large mass of iron in the ejecta of our models. The neutron-capture elements beyond zinc are produced by an \textsl{r}-process in neutron-rich jets. Their [X/Fe] abundances, however, are overproduced by a factor of 100 compared to those observed in \textsl{r}-process enhanced metal-poor stars. Models with jets that have a higher electron fraction ($Y_\mathrm{e} = 0.5$) can more naturally explain the chemical abundances observed in metal-poor stars without neutron-capture enhancement.

Thus, most of our findings support the magneto-rotational hypernova mechanism as a possible explanation for Ic-BL supernovae and for the abundances in extremely metal-poor stars. It is
noteworthy that we are able to obtain higher
nickel masses than the recent study of \citet{reichert_20} found for their 2D MHD simulations. Interestingly, though, we obtain
a very similar range of nickel masses $\mathord{\lesssim} 0.1\, \Msun$ with our calibrated jet models as they do in their MHD simulations for the same progenitor model 35OC. This lends confidence to the high nickel masses we find for progenitor model z25a. The key to producing the high nickel masses required to match observations might
lie in achieving similar conditions as in 
Models z25a-05 and z25a-10, i.e., not only
a powerful jet but also significant explosive burning
by the bow shock and efficient entrainment of
the nickel into the jet, which gives rises to
wider, lobe-like distribution of $^{56}\mathrm{Ni}$
in the ejecta. It will be worth investigating
in future how this can be achieved in more 
realistic models of magneto-rotational explosions,
e.g., by a favourable progenitor structure, or
by 3D effects that facilitate the entrainment of
material behind the bow shock.

On the other hand, the inability of our models
to reach high [C/Fe] poses a challenge for
the explanation of CEMP star abundances by magneto-rotational hypernovae. In fact, this predicament is largely independent of how the explosions are modelled and already comes from the observational constraints on the nickel masses of SNe Ic-BL and the progenitor structure, which limits the
amount of carbon that can be ejected. It will
need to be investigated further what the simultaneous
constraints from transients and CEMP stars imply
for the structure of hypernova progenitors.
As discussed in the introduction, it is possible
that other sources, e.g., faint Population III supernovae with considerable fallback might be responsible for CEMP  stars \citep{umeda_2005,tominaga_2007},
or that the high [C/Fe] abundances can be
explained by the intricacies of mixing in
the interstellar medium and star formation
 \citet{frebel_2007a,hartwig_2019}. It is
 also conceivable that a better determination
 of observational nickel masses from
 Ic-BL light curves may relieve the tension.
 If the light curves are (partly) magnetar-
 powered \citep[e.g.,][]{woosley_2010,greiner_2015,wang_2017,wang_2019}, the observational nickel 
 masses  could be overestimated. More detailed
 multi-dimensional radiative transfer modelling
 of SNe Ic-BL could also lead to revised estimates for the nickel mass. More rigorous modelling of hypernovae, their observational signatures, and their impact is clearly called for to better understand the nature of these powerful explosions.

\section*{Acknowledgements}
We thank the referee, Friedel Thielemann, for his constructive and thoughtful comments.
This work was supported by the Australian Research Council through ARC Future Fellowship FT160100035 (BM) and Future Fellowship FT120100363 (AH),  by the Australian Research Council Centre of Excellence for Gravitational Wave Discovery (OzGrav), through project number CE170100004; by the Australian Research Council Centre of Excellence for All Sky Astrophysics in 3 Dimensions (ASTRO 3D).
AH has been supported, in part, by a grant from Science and Technology Commission of Shanghai Municipality (Grants No.16DZ2260200) and National Natural Science Foundation of China (Grants No.11655002).
MO acknowledges support from the European Research Council under grant EUROPIUM-667912, from the the Deutsche Forschungsgemeinschaft (DFG, German Research Foundation) -- Projektnummer 279384907 -- SFB 1245 as well as from the and from the Spanish Ministry of Science, Education and Universities (PGC2018-095984-B-I00, RYC2018-024938-I) and the Valencian Community (PROMETEU/2019/071).
This material is based upon work supported by the National Science Foundation under Grant No. PHY-1430152 (JINA Center for the Evolution of the Elements).  This research was undertaken with the assistance of resources from the National Computational Infrastructure (NCI), which is supported by the Australian Government and was supported by resources provided by the Pawsey Supercomputing Centre with funding from the Australian Government and the Government of Western Australia. JG  acknowledges financial support from an Australian Government Research Training Program (RTP) Scholarship.

\section*{Data Availability}
The data underlying this article will be shared on reasonable request to the corresponding author, subject to considerations of intellectual property law.




\bibliographystyle{mnras}
\bibliography{james} 

\begin{thebibliography}{}
\makeatletter
\relax
\def\mn@urlcharsother{\let\do\@makeother \do\$\do\&\do\#\do\^\do\_\do\%\do\~}
\def\mn@doi{\begingroup\mn@urlcharsother \@ifnextchar [ {\mn@doi@}
  {\mn@doi@[]}}
\def\mn@doi@[#1]#2{\def\@tempa{#1}\ifx\@tempa\@empty \href
  {http://dx.doi.org/#2} {doi:#2}\else \href {http://dx.doi.org/#2} {#1}\fi
  \endgroup}
\def\mn@eprint#1#2{\mn@eprint@#1:#2::\@nil}
\def\mn@eprint@arXiv#1{\href {http://arxiv.org/abs/#1} {{\tt arXiv:#1}}}
\def\mn@eprint@dblp#1{\href {http://dblp.uni-trier.de/rec/bibtex/#1.xml}
  {dblp:#1}}
\def\mn@eprint@#1:#2:#3:#4\@nil{\def\@tempa {#1}\def\@tempb {#2}\def\@tempc
  {#3}\ifx \@tempc \@empty \let \@tempc \@tempb \let \@tempb \@tempa \fi \ifx
  \@tempb \@empty \def\@tempb {arXiv}\fi \@ifundefined
  {mn@eprint@\@tempb}{\@tempb:\@tempc}{\expandafter \expandafter \csname
  mn@eprint@\@tempb\endcsname \expandafter{\@tempc}}}

\bibitem[\protect\citeauthoryear{{Abohalima} \& {Frebel}}{{Abohalima} \&
  {Frebel}}{2018}]{abohalima_2018}
{Abohalima} A.,  {Frebel} A.,  2018, \mn@doi [\apjs]
  {10.3847/1538-4365/aadfe9}, \href
  {https://ui.adsabs.harvard.edu/abs/2018ApJS..238...36A} {238, 36}

\bibitem[\protect\citeauthoryear{{Aguilera-Dena}, {Langer}, {Moriya}  \&
  {Schootemeijer}}{{Aguilera-Dena} et~al.}{2018}]{aguilera_18}
{Aguilera-Dena} D.~R.,  {Langer} N.,  {Moriya} T.~J.,   {Schootemeijer} A.,
  2018, \mn@doi [\apj] {10.3847/1538-4357/aabfc1}, \href
  {https://ui.adsabs.harvard.edu/abs/2018ApJ...858..115A} {858, 115}

\bibitem[\protect\citeauthoryear{{Aguilera-Dena}, {Langer}, {Antoniadis}  \&
  {M{\"u}ller}}{{Aguilera-Dena} et~al.}{2020}]{aguilera_20}
{Aguilera-Dena} D.~R.,  {Langer} N.,  {Antoniadis} J.,   {M{\"u}ller} B.,
  2020, \mn@doi [\apj] {10.3847/1538-4357/abb138}, \href
  {https://ui.adsabs.harvard.edu/abs/2020ApJ...901..114A} {901, 114}

\bibitem[\protect\citeauthoryear{{Akiyama}, {Wheeler}, {Meier}  \&
  {Lichtenstadt}}{{Akiyama} et~al.}{2003}]{akiyama_2003}
{Akiyama} S.,  {Wheeler} J.~C.,  {Meier} D.~L.,   {Lichtenstadt} I.,  2003,
  \mn@doi [\apj] {10.1086/344135}, \href
  {http://adsabs.harvard.edu/abs/2003ApJ...584..954A} {584, 954}

\bibitem[\protect\citeauthoryear{{Anderson}}{{Anderson}}{2019}]{anderson_2019}
{Anderson} J.~P.,  2019, arXiv e-prints, \href
  {https://ui.adsabs.harvard.edu/abs/2019arXiv190600761A} {p. arXiv:1906.00761}

\bibitem[\protect\citeauthoryear{{Aoki}, {Beers}, {Christlieb}, {Norris},
  {Ryan}  \& {Tsangarides}}{{Aoki} et~al.}{2007}]{aoki_2007}
{Aoki} W.,  {Beers} T.~C.,  {Christlieb} N.,  {Norris} J.~E.,  {Ryan} S.~G.,
  {Tsangarides} S.,  2007, \mn@doi [\apj] {10.1086/509817}, \href
  {https://ui.adsabs.harvard.edu/abs/2007ApJ...655..492A} {655, 492}

\bibitem[\protect\citeauthoryear{{Arcavi} et~al.,}{{Arcavi}
  et~al.}{2010}]{arcavi_2010}
{Arcavi} I.,  et~al., 2010, \mn@doi [\apj] {10.1088/0004-637X/721/1/777}, \href
  {https://ui.adsabs.harvard.edu/abs/2010ApJ...721..777A} {721, 777}

\bibitem[\protect\citeauthoryear{{Arentsen}, {Starkenburg}, {Shetrone}, {Venn},
  {Depagne}  \& {McConnachie}}{{Arentsen} et~al.}{2019}]{arentsen_2019}
{Arentsen} A.,  {Starkenburg} E.,  {Shetrone} M.~D.,  {Venn} K.~A.,  {Depagne}
  {\'E}.,   {McConnachie} A.~W.,  2019, \mn@doi [\aap]
  {10.1051/0004-6361/201834146}, \href
  {https://ui.adsabs.harvard.edu/abs/2019A&A...621A.108A} {621, A108}

\bibitem[\protect\citeauthoryear{Ashall et~al.,}{Ashall
  et~al.}{2019}]{ashall_2019}
Ashall C.,  et~al., 2019, \mn@doi [Monthly Notices of the Royal Astronomical
  Society] {10.1093/mnras/stz1588}, 487, 5824–5839

\bibitem[\protect\citeauthoryear{{Banerjee}, {Heger}  \& {Qian}}{{Banerjee}
  et~al.}{2019}]{banerjee_2019}
{Banerjee} P.,  {Heger} A.,   {Qian} Y.-Z.,  2019, \mn@doi [\apj]
  {10.3847/1538-4357/ab517a}, \href
  {https://ui.adsabs.harvard.edu/abs/2019ApJ...887..187B} {887, 187}

\bibitem[\protect\citeauthoryear{{Barnes}, {Duffell}, {Liu}, {Modjaz},
  {Bianco}, {Kasen}  \& {MacFadyen}}{{Barnes} et~al.}{2018}]{barnes_2018}
{Barnes} J.,  {Duffell} P.~C.,  {Liu} Y.,  {Modjaz} M.,  {Bianco} F.~B.,
  {Kasen} D.,   {MacFadyen} A.~I.,  2018, \mn@doi [\apj]
  {10.3847/1538-4357/aabf84}, \href
  {https://ui.adsabs.harvard.edu/abs/2018ApJ...860...38B} {860, 38}

\bibitem[\protect\citeauthoryear{Beers \& Christlieb}{Beers \&
  Christlieb}{2005}]{beers_2005}
Beers T.~C.,  Christlieb N.,  2005, \mn@doi [\araa]
  {10.1146/annurev.astro.42.053102.134057}, 43, 531

\bibitem[\protect\citeauthoryear{{Blackman}, {Nordhaus}  \&
  {Thomas}}{{Blackman} et~al.}{2006}]{blackman_2006}
{Blackman} E.~G.,  {Nordhaus} J.~T.,   {Thomas} J.~H.,  2006, \mn@doi [\na]
  {10.1016/j.newast.2005.11.003}, \href
  {https://ui.adsabs.harvard.edu/\#abs/2006NewA...11..452B} {11, 452}

\bibitem[\protect\citeauthoryear{{Burrows}, {Dessart}, {Livne}, {Ott}  \&
  {Murphy}}{{Burrows} et~al.}{2007}]{burrows_2007}
{Burrows} A.,  {Dessart} L.,  {Livne} E.,  {Ott} C.~D.,   {Murphy} J.,  2007,
  \mn@doi [\apj] {10.1086/519161}, \href
  {http://adsabs.harvard.edu/abs/2007ApJ...664..416B} {664, 416}

\bibitem[\protect\citeauthoryear{{Burrows}, {Radice}, {Vartanyan}, {Nagakura},
  {Skinner}  \& {Dolence}}{{Burrows} et~al.}{2020}]{burrows_2020}
{Burrows} A.,  {Radice} D.,  {Vartanyan} D.,  {Nagakura} H.,  {Skinner} M.~A.,
   {Dolence} J.~C.,  2020, \mn@doi [\mnras] {10.1093/mnras/stz3223}, \href
  {https://ui.adsabs.harvard.edu/abs/2020MNRAS.491.2715B} {491, 2715}

\bibitem[\protect\citeauthoryear{{Cayrel} et~al.,}{{Cayrel}
  et~al.}{2004}]{cayrel_2004}
{Cayrel} R.,  et~al., 2004, \aap, \href
  {http://adsabs.harvard.edu/abs/2004A26A...416.1117C} {416, 1117}

\bibitem[\protect\citeauthoryear{{Chen}, {Moriya}, {Woosley}, {Sukhbold},
  {Whalen}, {Suwa}  \& {Bromm}}{{Chen} et~al.}{2017}]{chen_2017}
{Chen} K.-J.,  {Moriya} T.~J.,  {Woosley} S.,  {Sukhbold} T.,  {Whalen} D.~J.,
  {Suwa} Y.,   {Bromm} V.,  2017, \mn@doi [\apj] {10.3847/1538-4357/aa68a4},
  \href {https://ui.adsabs.harvard.edu/abs/2017ApJ...839...85C} {839, 85}

\bibitem[\protect\citeauthoryear{{Chiaki} \& {Wise}}{{Chiaki} \&
  {Wise}}{2019}]{chiaki_2019}
{Chiaki} G.,  {Wise} J.~H.,  2019, \mn@doi [\mnras] {10.1093/mnras/sty2984},
  \href {https://ui.adsabs.harvard.edu/abs/2019MNRAS.482.3933C} {482, 3933}

\bibitem[\protect\citeauthoryear{{Corsi} \& {M{\'e}sz{\'a}ros}}{{Corsi} \&
  {M{\'e}sz{\'a}ros}}{2009}]{corsi_09}
{Corsi} A.,  {M{\'e}sz{\'a}ros} P.,  2009, \mn@doi [\apj]
  {10.1088/0004-637X/702/2/1171}, \href
  {http://adsabs.harvard.edu/abs/2009ApJ...702.1171C} {702, 1171}

\bibitem[\protect\citeauthoryear{Cyburt et~al.,}{Cyburt
  et~al.}{2010}]{cyburt_2010}
Cyburt R.~H.,  et~al., 2010, \mn@doi [The Astrophysical Journal Supplement
  Series] {10.1088/0067-0049/189/1/240}, 189, 240

\bibitem[\protect\citeauthoryear{{Detmers}, {Langer}, {Podsiadlowski}  \&
  {Izzard}}{{Detmers} et~al.}{2008}]{detmers_08}
{Detmers} R.~G.,  {Langer} N.,  {Podsiadlowski} P.,   {Izzard} R.~G.,  2008,
  \mn@doi [\aap] {10.1051/0004-6361:200809371}, \href
  {https://ui.adsabs.harvard.edu/abs/2008A&A...484..831D} {484, 831}

\bibitem[\protect\citeauthoryear{{Dimmelmeier}, {Font}  \&
  {M{\"u}ller}}{{Dimmelmeier} et~al.}{2002}]{dimmelmeier_2002}
{Dimmelmeier} H.,  {Font} J.~A.,   {M{\"u}ller} E.,  2002, \mn@doi [\aap]
  {10.1051/0004-6361:20020563}, \href
  {https://ui.adsabs.harvard.edu/abs/2002A&A...388..917D} {388, 917}

\bibitem[\protect\citeauthoryear{Dimmelmeier, Novak, Font, Ib\'a\~nez  \&
  M\"uller}{Dimmelmeier et~al.}{2005}]{dimmelmeier_2005}
Dimmelmeier H.,  Novak J.,  Font J.~A.,  Ib\'a\~nez J.~M.,   M\"uller E.,
  2005, \mn@doi [Phys. Rev. D] {10.1103/PhysRevD.71.064023}, 71, 064023

\bibitem[\protect\citeauthoryear{{Duffell} \& {MacFadyen}}{{Duffell} \&
  {MacFadyen}}{2015}]{duffell_15}
{Duffell} P.~C.,  {MacFadyen} A.~I.,  2015, \mn@doi [\apj]
  {10.1088/0004-637X/806/2/205}, \href
  {http://adsabs.harvard.edu/abs/2015ApJ...806..205D} {806, 205}

\bibitem[\protect\citeauthoryear{Ezzeddine et~al.,}{Ezzeddine
  et~al.}{2019}]{ezzeddine_2019}
Ezzeddine R.,  et~al., 2019, \mn@doi [The Astrophysical Journal]
  {10.3847/1538-4357/ab14e7}, 876, 97

\bibitem[\protect\citeauthoryear{Frebel, Johnson  \& Bromm}{Frebel
  et~al.}{2007}]{frebel_2007a}
Frebel A.,  Johnson J.~L.,   Bromm V.,  2007, \mn@doi [Monthly Notices of the
  Royal Astronomical Society: Letters] {10.1111/j.1745-3933.2007.00344.x}, 380,
  L40–L44

\bibitem[\protect\citeauthoryear{{Fujimoto}, {Hashimoto}, {Kotake}  \&
  {Yamada}}{{Fujimoto} et~al.}{2007}]{fujimoto_2007}
{Fujimoto} S.,  {Hashimoto} M.,  {Kotake} K.,   {Yamada} S.,  2007, \apj, 656,
  382

\bibitem[\protect\citeauthoryear{Gehrels, Ramirez-Ruiz  \& Fox}{Gehrels
  et~al.}{2009}]{gehrels_2009}
Gehrels N.,  Ramirez-Ruiz E.,   Fox D.,  2009, \mn@doi [Annual Review of
  Astronomy and Astrophysics] {10.1146/annurev.astro.46.060407.145147}, 47, 567

\bibitem[\protect\citeauthoryear{{Gompertz}, {O'Brien}  \& {Wynn}}{{Gompertz}
  et~al.}{2014}]{gompertz_14}
{Gompertz} B.~P.,  {O'Brien} P.~T.,   {Wynn} G.~A.,  2014, \mn@doi [\mnras]
  {10.1093/mnras/stt2165}, \href
  {http://adsabs.harvard.edu/abs/2014MNRAS.438..240G} {438, 240}

\bibitem[\protect\citeauthoryear{{Greiner} et~al.,}{{Greiner}
  et~al.}{2015}]{greiner_2015}
{Greiner} J.,  et~al., 2015, \mn@doi [\nat] {10.1038/nature14579}, \href
  {https://ui.adsabs.harvard.edu/abs/2015Natur.523..189G} {523, 189}

\bibitem[\protect\citeauthoryear{{Halevi} \& {M{\"o}sta}}{{Halevi} \&
  {M{\"o}sta}}{2018}]{halevi_2018}
{Halevi} G.,  {M{\"o}sta} P.,  2018, \mn@doi [\mnras] {10.1093/mnras/sty797},
  \href {http://adsabs.harvard.edu/abs/2018MNRAS.477.2366H} {477, 2366}

\bibitem[\protect\citeauthoryear{{Hamann} \& {Koesterke}}{{Hamann} \&
  {Koesterke}}{1998}]{hamann_1998}
{Hamann} W.-R.,  {Koesterke} L.,  1998, \aap, \href
  {http://adsabs.harvard.edu/abs/1998A\%26A...335.1003H} {335, 1003}

\bibitem[\protect\citeauthoryear{{Hamann}, {Koesterke}  \&
  {Wessolowski}}{{Hamann} et~al.}{1995}]{hamann_1995}
{Hamann} W.-R.,  {Koesterke} L.,   {Wessolowski} U.,  1995, \aap, \href
  {http://adsabs.harvard.edu/abs/1995A%26A...299..151H} {299, 151}

\bibitem[\protect\citeauthoryear{{Hartwig} \& {Yoshida}}{{Hartwig} \&
  {Yoshida}}{2019}]{hartwig_2019}
{Hartwig} T.,  {Yoshida} N.,  2019, \mn@doi [\apjl] {10.3847/2041-8213/aaf866},
  \href {https://ui.adsabs.harvard.edu/abs/2019ApJ...870L...3H} {870, L3}

\bibitem[\protect\citeauthoryear{{Heger}, {Langer}  \& {Woosley}}{{Heger}
  et~al.}{2000}]{heger_2000}
{Heger} A.,  {Langer} N.,   {Woosley} S.~E.,  2000, \mn@doi [\apj]
  {10.1086/308158}, \href
  {https://ui.adsabs.harvard.edu/abs/2000ApJ...528..368H} {528, 368}

\bibitem[\protect\citeauthoryear{{Heger}, {Woosley}  \& {Spruit}}{{Heger}
  et~al.}{2005}]{heger_2005}
{Heger} A.,  {Woosley} S.~E.,   {Spruit} H.~C.,  2005, \mn@doi [\apj]
  {10.1086/429868}, \href {http://adsabs.harvard.edu/abs/2005ApJ...626..350H}
  {626, 350}

\bibitem[\protect\citeauthoryear{{Iwamoto} et~al.,}{{Iwamoto}
  et~al.}{1998}]{iwamoto_1998}
{Iwamoto} K.,  et~al., 1998, Nature, \href
  {http://adsabs.harvard.edu/abs/1998Natur.395..672I} {395, 672}

\bibitem[\protect\citeauthoryear{{Janka}}{{Janka}}{2017}]{janka_2017}
{Janka} H.-T.,  2017, {Neutrino-Driven Explosions}.
p.~1095, \mn@doi{10.1007/978-3-319-21846-5_109}

\bibitem[\protect\citeauthoryear{{Ji}, {Frebel}  \& {Bromm}}{{Ji}
  et~al.}{2015}]{ji_2015}
{Ji} A.~P.,  {Frebel} A.,   {Bromm} V.,  2015, \mn@doi [\mnras]
  {10.1093/mnras/stv2052}, \href
  {https://ui.adsabs.harvard.edu/abs/2015MNRAS.454..659J} {454, 659}

\bibitem[\protect\citeauthoryear{{Kobayashi}, {Umeda}, {Nomoto}, {Tominaga}  \&
  {Ohkubo}}{{Kobayashi} et~al.}{2006}]{kobayashi_2006}
{Kobayashi} C.,  {Umeda} H.,  {Nomoto} K.,  {Tominaga} N.,   {Ohkubo} T.,
  2006, \apj, \href {http://adsabs.harvard.edu/abs/2006ApJ...653.1145K} {653,
  1145}

\bibitem[\protect\citeauthoryear{{Kobayashi}, {Karakas}  \&
  {Lugaro}}{{Kobayashi} et~al.}{2020}]{kobayashi_2020}
{Kobayashi} C.,  {Karakas} A.~I.,   {Lugaro} M.,  2020, \mn@doi [\apj]
  {10.3847/1538-4357/abae65}, \href
  {https://ui.adsabs.harvard.edu/abs/2020ApJ...900..179K} {900, 179}

\bibitem[\protect\citeauthoryear{Komiya, Suda, Yamada  \& Fujimoto}{Komiya
  et~al.}{2020}]{komiya_2020}
Komiya Y.,  Suda T.,  Yamada S.,   Fujimoto M.~Y.,  2020, \mn@doi [The
  Astrophysical Journal] {10.3847/1538-4357/ab67be}, 890, 66

\bibitem[\protect\citeauthoryear{{Kuroda}, {Arcones}, {Takiwaki}  \&
  {Kotake}}{{Kuroda} et~al.}{2020}]{kuroda_2020}
{Kuroda} T.,  {Arcones} A.,  {Takiwaki} T.,   {Kotake} K.,  2020, \mn@doi
  [\apj] {10.3847/1538-4357/ab9308}, \href
  {https://ui.adsabs.harvard.edu/abs/2020ApJ...896..102K} {896, 102}

\bibitem[\protect\citeauthoryear{{Lattimer} \& {Swesty}}{{Lattimer} \&
  {Swesty}}{1991}]{lattimer_1991}
{Lattimer} J.~M.,  {Swesty} D.~F.,  1991, \mn@doi [\nphysa]
  {10.1016/0375-9474(91)90452-C}, \href
  {https://ui.adsabs.harvard.edu/abs/1991NuPhA.535..331L} {535, 331}

\bibitem[\protect\citeauthoryear{{Lippuner} \& {Roberts}}{{Lippuner} \&
  {Roberts}}{2017}]{lippuner_2017}
{Lippuner} J.,  {Roberts} L.~F.,  2017, \mn@doi [\apjs]
  {10.3847/1538-4365/aa94cb}, \href
  {https://ui.adsabs.harvard.edu/abs/2017ApJS..233...18L} {233, 18}

\bibitem[\protect\citeauthoryear{{MacFadyen} \& {Woosley}}{{MacFadyen} \&
  {Woosley}}{1999}]{macfadyen_1999}
{MacFadyen} A.~I.,  {Woosley} S.~E.,  1999, \mn@doi [\apj] {10.1086/307790},
  \href {http://adsabs.harvard.edu/abs/1999ApJ...524..262M} {524, 262}

\bibitem[\protect\citeauthoryear{{MacFadyen}, {Woosley}  \&
  {Heger}}{{MacFadyen} et~al.}{2001}]{macfadyen_2001}
{MacFadyen} A.~I.,  {Woosley} S.~E.,   {Heger} A.,  2001, \apj, \href
  {http://adsabs.harvard.edu/abs/2001ApJ...550..410M} {550, 410}

\bibitem[\protect\citeauthoryear{{Maeda} \& {Nomoto}}{{Maeda} \&
  {Nomoto}}{2003}]{maeda_2003}
{Maeda} K.,  {Nomoto} K.,  2003, \mn@doi [\apj] {10.1086/378948}, \href
  {http://adsabs.harvard.edu/abs/2003ApJ...598.1163M} {598, 1163}

\bibitem[\protect\citeauthoryear{Maeda, Nakamura, Nomoto, Mazzali, Patat  \&
  Hachisu}{Maeda et~al.}{2002}]{maeda_2002}
Maeda K.,  Nakamura T.,  Nomoto K.,  Mazzali P.~A.,  Patat F.,   Hachisu I.,
  2002, \apj, 565, 405

\bibitem[\protect\citeauthoryear{Maeda et~al.,}{Maeda
  et~al.}{2008}]{maeda_2008}
Maeda K.,  et~al., 2008, \mn@doi [Science] {10.1126/science.1149437}, 319, 1220

\bibitem[\protect\citeauthoryear{{Maeder}, {Meynet}  \& {Chiappini}}{{Maeder}
  et~al.}{2015}]{maeder_2015}
{Maeder} A.,  {Meynet} G.,   {Chiappini} C.,  2015, \mn@doi [\aap]
  {10.1051/0004-6361/201424153}, \href
  {http://adsabs.harvard.edu/abs/2015A\%26A...576A..56M} {576, A56}

\bibitem[\protect\citeauthoryear{{Magg} et~al.,}{{Magg}
  et~al.}{2020}]{magg_2020}
{Magg} M.,  et~al., 2020, \mn@doi [\mnras] {10.1093/mnras/staa2624}, \href
  {https://ui.adsabs.harvard.edu/abs/2020MNRAS.tmp.2008M} {}

\bibitem[\protect\citeauthoryear{{Mazzali}, {Nomoto}, {Patat}  \&
  {Maeda}}{{Mazzali} et~al.}{2001}]{mazzali_2001}
{Mazzali} P.~A.,  {Nomoto} K.,  {Patat} F.,   {Maeda} K.,  2001, \mn@doi [\apj]
  {10.1086/322420}, \href
  {https://ui.adsabs.harvard.edu/abs/2001ApJ...559.1047M} {559, 1047}

\bibitem[\protect\citeauthoryear{{Mazzali}, {McFadyen}, {Woosley}, {Pian}  \&
  {Tanaka}}{{Mazzali} et~al.}{2014}]{mazzali_2014}
{Mazzali} P.~A.,  {McFadyen} A.~I.,  {Woosley} S.~E.,  {Pian} E.,   {Tanaka}
  M.,  2014, \mn@doi [\mnras] {10.1093/mnras/stu1124}, \href
  {http://adsabs.harvard.edu/abs/2014MNRAS.443...67M} {443, 67}

\bibitem[\protect\citeauthoryear{{McWilliam}, {Preston}, {Sneden}  \&
  {Shectman}}{{McWilliam} et~al.}{1995a}]{mcwilliam_1995a}
{McWilliam} A.,  {Preston} G.~W.,  {Sneden} C.,   {Shectman} S.,  1995a,
  \mn@doi [\aj] {10.1086/117485}, \href
  {http://adsabs.harvard.edu/abs/1995AJ....109.2736M} {109, 2736}

\bibitem[\protect\citeauthoryear{{McWilliam}, {Preston}, {Sneden}  \&
  {Searle}}{{McWilliam} et~al.}{1995b}]{mcwilliam_1995b}
{McWilliam} A.,  {Preston} G.~W.,  {Sneden} C.,   {Searle} L.,  1995b, \mn@doi
  [\aj] {10.1086/117486}, \href
  {http://adsabs.harvard.edu/abs/1995AJ....109.2757M} {109, 2757}

\bibitem[\protect\citeauthoryear{{Meynet}, {Ekstr{\"o}m}  \& {Maeder}}{{Meynet}
  et~al.}{2006}]{meynet_2006}
{Meynet} G.,  {Ekstr{\"o}m} S.,   {Maeder} A.,  2006, \mn@doi [Astronomy \&
  Astrophysics] {10.1051/0004-6361:20053070}, \href
  {http://adsabs.harvard.edu/abs/2006A%26A...447..623M} {447, 623}

\bibitem[\protect\citeauthoryear{{Modjaz}, {Liu}, {Bianco}  \&
  {Graur}}{{Modjaz} et~al.}{2016}]{modjaz_2016}
{Modjaz} M.,  {Liu} Y.~Q.,  {Bianco} F.~B.,   {Graur} O.,  2016, \mn@doi [\apj]
  {10.3847/0004-637X/832/2/108}, \href
  {https://ui.adsabs.harvard.edu/abs/2016ApJ...832..108M} {832, 108}

\bibitem[\protect\citeauthoryear{{M{\"o}sta} et~al.,}{{M{\"o}sta}
  et~al.}{2014}]{mosta_2014}
{M{\"o}sta} P.,  et~al., 2014, \mn@doi [\apjl] {10.1088/2041-8205/785/2/L29},
  \href {http://adsabs.harvard.edu/abs/2014ApJ...785L..29M} {785, L29}

\bibitem[\protect\citeauthoryear{{M{\"o}sta}, {Ott}, {Radice}, {Roberts},
  {Schnetter}  \& {Haas}}{{M{\"o}sta} et~al.}{2015}]{mosta_2015}
{M{\"o}sta} P.,  {Ott} C.~D.,  {Radice} D.,  {Roberts} L.~F.,  {Schnetter} E.,
   {Haas} R.,  2015, \mn@doi [\nat] {10.1038/nature15755}, \href
  {http://adsabs.harvard.edu/abs/2015Natur.528..376M} {528, 376}

\bibitem[\protect\citeauthoryear{M{\"o}sta, Roberts, Halevi, Ott, Lippuner,
  Haas  \& Schnetter}{M{\"o}sta et~al.}{2018a}]{mosta_2018}
M{\"o}sta P.,  Roberts L.~F.,  Halevi G.,  Ott C.~D.,  Lippuner J.,  Haas R.,
  Schnetter E.,  2018a, \mn@doi [The Astrophysical Journal]
  {10.3847/1538-4357/aad6ec}, 864, 171

\bibitem[\protect\citeauthoryear{{M{\"o}sta}, {Roberts}, {Halevi}, {Ott},
  {Lippuner}, {Haas}  \& {Schnetter}}{{M{\"o}sta} et~al.}{2018b}]{moesta_2018}
{M{\"o}sta} P.,  {Roberts} L.~F.,  {Halevi} G.,  {Ott} C.~D.,  {Lippuner} J.,
  {Haas} R.,   {Schnetter} E.,  2018b, \mn@doi [\apj]
  {10.3847/1538-4357/aad6ec}, \href
  {https://ui.adsabs.harvard.edu/abs/2018ApJ...864..171M} {864, 171}

\bibitem[\protect\citeauthoryear{{M{\"u}ller}}{{M{\"u}ller}}{2016}]{muller_2016}
{M{\"u}ller} B.,  2016, \mn@doi [\pasa] {10.1017/pasa.2016.40}, \href
  {http://adsabs.harvard.edu/abs/2016PASA...33...48M} {33, e048}

\bibitem[\protect\citeauthoryear{{M{\"u}ller} \& {Janka}}{{M{\"u}ller} \&
  {Janka}}{2015}]{mueller_2015}
{M{\"u}ller} B.,  {Janka} H.~T.,  2015, \mn@doi [\mnras]
  {10.1093/mnras/stv101}, \href
  {https://ui.adsabs.harvard.edu/abs/2015MNRAS.448.2141M} {448, 2141}

\bibitem[\protect\citeauthoryear{{M{\"u}ller}, {Janka}  \&
  {Dimmelmeier}}{{M{\"u}ller} et~al.}{2010}]{mueller_2010}
{M{\"u}ller} B.,  {Janka} H.-T.,   {Dimmelmeier} H.,  2010, \mn@doi [\apjs]
  {10.1088/0067-0049/189/1/104}, \href
  {http://adsabs.harvard.edu/abs/2010ApJS..189..104M} {189, 104}

\bibitem[\protect\citeauthoryear{{M{\"u}ller}, {Melson}, {Heger}  \&
  {Janka}}{{M{\"u}ller} et~al.}{2017}]{muller_2017}
{M{\"u}ller} B.,  {Melson} T.,  {Heger} A.,   {Janka} H.-T.,  2017, \mn@doi
  [\mnras] {10.1093/mnras/stx1962}, \href
  {https://ui.adsabs.harvard.edu/abs/2017MNRAS.472..491M} {472, 491}

\bibitem[\protect\citeauthoryear{Nagataki, Mizuta  \& Sato}{Nagataki
  et~al.}{2006}]{nagataki_2006}
Nagataki S.,  Mizuta A.,   Sato K.,  2006, \apj, 647, 1255

\bibitem[\protect\citeauthoryear{{Nakamura}, {Mazzali}, {Nomoto}  \&
  {Iwamoto}}{{Nakamura} et~al.}{2001a}]{nakamura_2001a}
{Nakamura} T.,  {Mazzali} P.~A.,  {Nomoto} K.,   {Iwamoto} K.,  2001a, \mn@doi
  [\apj] {10.1086/319784}, \href
  {https://ui.adsabs.harvard.edu/abs/2001ApJ...550..991N} {550, 991}

\bibitem[\protect\citeauthoryear{{Nakamura}, {Umeda}, {Iwamoto}, {Nomoto},
  {Hashimoto}, {Hix}  \& {Thielemann}}{{Nakamura}
  et~al.}{2001b}]{nakamura_2001}
{Nakamura} T.,  {Umeda} H.,  {Iwamoto} K.,  {Nomoto} K.,  {Hashimoto} M.-a.,
  {Hix} W.~R.,   {Thielemann} F.-K.,  2001b, \apj, \href
  {http://adsabs.harvard.edu/abs/2001ApJ...555..880N} {555, 880}

\bibitem[\protect\citeauthoryear{{Nakamura}, {Kajino}, {Mathews}, {Sato}  \&
  {Harikae}}{{Nakamura} et~al.}{2015}]{nakamura_2015}
{Nakamura} K.,  {Kajino} T.,  {Mathews} G.~J.,  {Sato} S.,   {Harikae} S.,
  2015, \mn@doi [\aap] {10.1051/0004-6361/201526110}, \href
  {http://adsabs.harvard.edu/abs/2015A%26A...582A..34N} {582, A34}

\bibitem[\protect\citeauthoryear{{Nieuwenhuijzen} \& {de
  Jager}}{{Nieuwenhuijzen} \& {de Jager}}{1990}]{nieu_1990}
{Nieuwenhuijzen} H.,  {de Jager} C.,  1990, \aap, \href
  {http://adsabs.harvard.edu/abs/1990A%26A...231..134N} {231, 134}

\bibitem[\protect\citeauthoryear{{Nishimura}, {Kotake}, {Hashimoto}, {Yamada},
  {Nishimura}, {Fujimoto}  \& {Sato}}{{Nishimura}
  et~al.}{2006}]{nishimura_2006}
{Nishimura} S.,  {Kotake} K.,  {Hashimoto} M.-a.,  {Yamada} S.,  {Nishimura}
  N.,  {Fujimoto} S.,   {Sato} K.,  2006, \mn@doi [\apj] {10.1086/500786},
  \href {https://ui.adsabs.harvard.edu/abs/2006ApJ...642..410N} {642, 410}

\bibitem[\protect\citeauthoryear{{Nishimura}, {Takiwaki}  \&
  {Thielemann}}{{Nishimura} et~al.}{2015}]{nishimura_2015}
{Nishimura} N.,  {Takiwaki} T.,   {Thielemann} F.-K.,  2015, \mn@doi [\apj]
  {10.1088/0004-637X/810/2/109}, \href
  {https://ui.adsabs.harvard.edu/abs/2015ApJ...810..109N} {810, 109}

\bibitem[\protect\citeauthoryear{{Nishimura}, {Sawai}, {Takiwaki}, {Yamada}  \&
  {Thielemann}}{{Nishimura} et~al.}{2017}]{nishimura_2017}
{Nishimura} N.,  {Sawai} H.,  {Takiwaki} T.,  {Yamada} S.,   {Thielemann}
  F.-K.,  2017, \mn@doi [\apjl] {10.3847/2041-8213/aa5dee}, \href
  {http://adsabs.harvard.edu/abs/2017ApJ...836L..21N} {836, L21}

\bibitem[\protect\citeauthoryear{{Nomoto}, {Kobayashi}  \& {Tominaga}}{{Nomoto}
  et~al.}{2013}]{nomoto_2013}
{Nomoto} K.,  {Kobayashi} C.,   {Tominaga} N.,  2013, \araa, \href
  {http://adsabs.harvard.edu/abs/2013ARA\%26A..51..457N} {51, 457}

\bibitem[\protect\citeauthoryear{{Obergaulinger} \& {Aloy}}{{Obergaulinger} \&
  {Aloy}}{2017}]{obergaulinger_2017}
{Obergaulinger} M.,  {Aloy} M.~{\'A}.,  2017, \mn@doi [\mnras]
  {10.1093/mnrasl/slx046}, \href
  {http://adsabs.harvard.edu/abs/2017MNRAS.469L..43O} {469, L43}

\bibitem[\protect\citeauthoryear{Obergaulinger \& Aloy}{Obergaulinger \&
  Aloy}{2020}]{obergaulinger_2020}
Obergaulinger M.,  Aloy M.~A.,  2020, \mn@doi [Monthly Notices of the Royal
  Astronomical Society] {10.1093/mnras/staa096}, 492, 4613–4634

\bibitem[\protect\citeauthoryear{{Obergaulinger}, {Just}  \&
  {Aloy}}{{Obergaulinger} et~al.}{2018}]{obergaulinger_2018}
{Obergaulinger} M.,  {Just} O.,   {Aloy} M.~A.,  2018, \mn@doi [Journal of
  Physics G Nuclear Physics] {10.1088/1361-6471/aac982}, \href
  {https://ui.adsabs.harvard.edu/abs/2018JPhG...45h4001O} {45, 084001}

\bibitem[\protect\citeauthoryear{Ono, Hashimoto, Fujimoto, Kotake  \&
  Yamada}{Ono et~al.}{2009}]{ono_2009}
Ono M.,  Hashimoto M.-a.,  Fujimoto S.-i.,  Kotake K.,   Yamada S.,  2009,
  \mn@doi [Progress of Theoretical Physics] {10.1143/PTP.122.755}, 122, 755

\bibitem[\protect\citeauthoryear{{Ono}, {Hashimoto}, {Fujimoto}, {Kotake}  \&
  {Yamada}}{{Ono} et~al.}{2012}]{ono_2012}
{Ono} M.,  {Hashimoto} M.,  {Fujimoto} S.,  {Kotake} K.,   {Yamada} S.,  2012,
  \mn@doi [Progress of Theoretical Physics] {10.1143/PTP.128.741}, \href
  {https://ui.adsabs.harvard.edu/abs/2012PThPh.128..741O} {128, 741}

\bibitem[\protect\citeauthoryear{{Paczy{\'n}ski}}{{Paczy{\'n}ski}}{1998}]{paczynski_1998}
{Paczy{\'n}ski} B.,  1998, \mn@doi [\apjl] {10.1086/311148}, \href
  {http://adsabs.harvard.edu/abs/1998ApJ...494L..45P} {494, L45}

\bibitem[\protect\citeauthoryear{Placco, Frebel, Beers  \& Stancliffe}{Placco
  et~al.}{2014}]{placco_2014}
Placco V.~M.,  Frebel A.,  Beers T.~C.,   Stancliffe R.~J.,  2014, \mn@doi [The
  Astrophysical Journal] {10.1088/0004-637x/797/1/21}, 797, 21

\bibitem[\protect\citeauthoryear{Podsiadlowski, Mazzali, Nomoto, Lazzati  \&
  Cappellaro}{Podsiadlowski et~al.}{2004}]{podsiadlowski_2004}
Podsiadlowski P.,  Mazzali P.~A.,  Nomoto K.,  Lazzati D.,   Cappellaro E.,
  2004, \apjl, 607, L17

\bibitem[\protect\citeauthoryear{{Raynaud}, {Guilet}, {Janka}  \&
  {Gastine}}{{Raynaud} et~al.}{2020}]{raynaud_2020}
{Raynaud} R.,  {Guilet} J.,  {Janka} H.-T.,   {Gastine} T.,  2020, \mn@doi
  [Science Advances] {10.1126/sciadv.aay2732}, \href
  {https://ui.adsabs.harvard.edu/abs/2020SciA....6.2732R} {6, eaay2732}

\bibitem[\protect\citeauthoryear{{Reichert}, {Obergaulinger}, {Eichler}, {Aloy}
   \& {Arcones}}{{Reichert} et~al.}{2020}]{reichert_20}
{Reichert} M.,  {Obergaulinger} M.,  {Eichler} M.,  {Aloy} M.-{\'A}.,
  {Arcones} A.,  2020, arXiv e-prints, \href
  {https://ui.adsabs.harvard.edu/abs/2020arXiv201002227R} {p. arXiv:2010.02227}

\bibitem[\protect\citeauthoryear{Ritter, Sluder, Safranek-Shrader,
  Milosavljevic  \& Bromm}{Ritter et~al.}{2015}]{ritter_2015}
Ritter J.~S.,  Sluder A.,  Safranek-Shrader C.,  Milosavljevic M.,   Bromm V.,
  2015, \mn@doi [Monthly Notices of the Royal Astronomical Society]
  {10.1093/mnras/stv982}, 451, 1190

\bibitem[\protect\citeauthoryear{{Ryan}, {Norris}  \& {Beers}}{{Ryan}
  et~al.}{1996}]{ryan_1996}
{Ryan} S.~G.,  {Norris} J.~E.,   {Beers} T.~C.,  1996, \mn@doi [\apj]
  {10.1086/177967}, \href {http://adsabs.harvard.edu/abs/1996ApJ...471..254R}
  {471, 254}

\bibitem[\protect\citeauthoryear{Smidt, Whalen, Wiggins, Even, Johnson  \&
  Fryer}{Smidt et~al.}{2014}]{smidt_2014}
Smidt J.,  Whalen D.~J.,  Wiggins B.~K.,  Even W.,  Johnson J.~L.,   Fryer
  C.~L.,  2014, \apj, 797, 97

\bibitem[\protect\citeauthoryear{{Smith}, {Li}, {Filippenko}  \&
  {Chornock}}{{Smith} et~al.}{2011}]{smith_2011}
{Smith} N.,  {Li} W.,  {Filippenko} A.~V.,   {Chornock} R.,  2011, \mn@doi
  [\mnras] {10.1111/j.1365-2966.2011.17229.x}, \href
  {https://ui.adsabs.harvard.edu/abs/2011MNRAS.412.1522S} {412, 1522}

\bibitem[\protect\citeauthoryear{{Spruit}}{{Spruit}}{2002}]{spruit_2002}
{Spruit} H.~C.,  2002, \mn@doi [\aap] {10.1051/0004-6361:20011465}, \href
  {https://ui.adsabs.harvard.edu/abs/2002A&A...381..923S} {381, 923}

\bibitem[\protect\citeauthoryear{Stevance et~al.,}{Stevance
  et~al.}{2017}]{stevance_2017}
Stevance H.~F.,  et~al., 2017, \mn@doi [Monthly Notices of the Royal
  Astronomical Society] {10.1093/mnras/stx970}, 469, 1897

\bibitem[\protect\citeauthoryear{{Suwa} \& {Tominaga}}{{Suwa} \&
  {Tominaga}}{2015}]{suwa_2015}
{Suwa} Y.,  {Tominaga} N.,  2015, \mn@doi [\mnras] {10.1093/mnras/stv901},
  \href {https://ui.adsabs.harvard.edu/abs/2015MNRAS.451..282S} {451, 282}

\bibitem[\protect\citeauthoryear{{Taddia} et~al.,}{{Taddia}
  et~al.}{2019}]{taddia_2019}
{Taddia} F.,  et~al., 2019, \mn@doi [\aap] {10.1051/0004-6361/201834429}, \href
  {https://ui.adsabs.harvard.edu/abs/2019A&A...621A..71T} {621, A71}

\bibitem[\protect\citeauthoryear{{Tanaka}, {Maeda}, {Mazzali}, {Kawabata}  \&
  {Nomoto}}{{Tanaka} et~al.}{2017}]{tanaka_2017}
{Tanaka} M.,  {Maeda} K.,  {Mazzali} P.~A.,  {Kawabata} K.~S.,   {Nomoto} K.,
  2017, \mn@doi [\apj] {10.3847/1538-4357/aa6035}, \href
  {https://ui.adsabs.harvard.edu/abs/2017ApJ...837..105T} {837, 105}

\bibitem[\protect\citeauthoryear{{Thielemann}, {Nomoto}  \&
  {Hashimoto}}{{Thielemann} et~al.}{1996}]{thielemann_1996}
{Thielemann} F.-K.,  {Nomoto} K.,   {Hashimoto} M.-A.,  1996, \mn@doi [\apj]
  {10.1086/176980}, \href
  {https://ui.adsabs.harvard.edu/abs/1996ApJ...460..408T} {460, 408}

\bibitem[\protect\citeauthoryear{Thompson, Quataert  \& Burrows}{Thompson
  et~al.}{2005}]{thompson_2005}
Thompson T.~A.,  Quataert E.,   Burrows A.,  2005, \mn@doi [The Astrophysical
  Journal] {10.1086/427177}, 620, 861

\bibitem[\protect\citeauthoryear{{Tominaga}}{{Tominaga}}{2009}]{tominaga_2009}
{Tominaga} N.,  2009, \apjl, \href
  {http://adsabs.harvard.edu/abs/2009ApJ...690..526T} {690, 526}

\bibitem[\protect\citeauthoryear{{Tominaga}, {Umeda}  \& {Nomoto}}{{Tominaga}
  et~al.}{2007}]{tominaga_2007}
{Tominaga} N.,  {Umeda} H.,   {Nomoto} K.,  2007, \apj, \href
  {http://adsabs.harvard.edu/abs/2007ApJ...660..516T} {660, 516}

\bibitem[\protect\citeauthoryear{{Umeda} \& {Nomoto}}{{Umeda} \&
  {Nomoto}}{2002}]{umeda_2002}
{Umeda} H.,  {Nomoto} K.,  2002, \apj, \href
  {http://adsabs.harvard.edu/abs/2002ApJ...565..385U} {565, 385}

\bibitem[\protect\citeauthoryear{{Umeda} \& {Nomoto}}{{Umeda} \&
  {Nomoto}}{2005}]{umeda_2005}
{Umeda} H.,  {Nomoto} K.,  2005, \apj, \href
  {http://adsabs.harvard.edu/abs/2005ApJ...619..427U} {619, 427}

\bibitem[\protect\citeauthoryear{Wanajo, Müller, Janka  \& Heger}{Wanajo
  et~al.}{2018}]{wanajo_2018}
Wanajo S.,  Müller B.,  Janka H.-T.,   Heger A.,  2018, \mn@doi [\apj]
  {10.3847/1538-4357/aa9d97}, 852, 40

\bibitem[\protect\citeauthoryear{{Wang} \& {Wheeler}}{{Wang} \&
  {Wheeler}}{2008}]{wang_2008}
{Wang} L.,  {Wheeler} J.~C.,  2008, \mn@doi [\araa]
  {10.1146/annurev.astro.46.060407.145139}, \href
  {http://adsabs.harvard.edu/abs/2008ARA%26A..46..433W} {46, 433}

\bibitem[\protect\citeauthoryear{{Wang} et~al.,}{{Wang}
  et~al.}{2017}]{wang_2017}
{Wang} L.~J.,  et~al., 2017, \mn@doi [\apj] {10.3847/1538-4357/aa5ff5}, \href
  {https://ui.adsabs.harvard.edu/abs/2017ApJ...837..128W} {837, 128}

\bibitem[\protect\citeauthoryear{{Wang} et~al.,}{{Wang}
  et~al.}{2019}]{wang_2019}
{Wang} L.~J.,  et~al., 2019, \mn@doi [\mnras] {10.1093/mnras/stz2184}, \href
  {https://ui.adsabs.harvard.edu/abs/2019MNRAS.489.1110W} {489, 1110}

\bibitem[\protect\citeauthoryear{{Weaver}, {Zimmerman}  \& {Woosley}}{{Weaver}
  et~al.}{1978}]{weaver_1978}
{Weaver} T.~A.,  {Zimmerman} G.~B.,   {Woosley} S.~E.,  1978, \apj, \href
  {http://adsabs.harvard.edu/abs/1978ApJ...225.1021W} {225, 1021}

\bibitem[\protect\citeauthoryear{{Wheeler}, {Meier}  \& {Wilson}}{{Wheeler}
  et~al.}{2002}]{wheeler_2002}
{Wheeler} J.~C.,  {Meier} D.~L.,   {Wilson} J.~R.,  2002, \mn@doi [\apj]
  {10.1086/338953}, \href {http://adsabs.harvard.edu/abs/2002ApJ...568..807W}
  {568, 807}

\bibitem[\protect\citeauthoryear{{Winteler}, {K{\"a}ppeli}, {Perego},
  {Arcones}, {Vasset}, {Nishimura}, {Liebend{\"o}rfer}  \&
  {Thielemann}}{{Winteler} et~al.}{2012}]{winteler_2012}
{Winteler} C.,  {K{\"a}ppeli} R.,  {Perego} A.,  {Arcones} A.,  {Vasset} N.,
  {Nishimura} N.,  {Liebend{\"o}rfer} M.,   {Thielemann} F.-K.,  2012, \mn@doi
  [\apjl] {10.1088/2041-8205/750/1/L22}, \href
  {http://adsabs.harvard.edu/abs/2012ApJ...750L..22W} {750, L22}

\bibitem[\protect\citeauthoryear{{Woosley}}{{Woosley}}{2010}]{woosley_2010}
{Woosley} S.~E.,  2010, \mn@doi [\apjl] {10.1088/2041-8205/719/2/L204}, \href
  {https://ui.adsabs.harvard.edu/abs/2010ApJ...719L.204W} {719, L204}

\bibitem[\protect\citeauthoryear{{Woosley}}{{Woosley}}{2011}]{woosley_2011}
{Woosley} S.~E.,  2011, arXiv e-prints, \href
  {https://ui.adsabs.harvard.edu/\#abs/2011arXiv1105.4193W} {p.
  arXiv:1105.4193}

\bibitem[\protect\citeauthoryear{{Woosley} \& {Bloom}}{{Woosley} \&
  {Bloom}}{2006}]{woosley_2006}
{Woosley} S.~E.,  {Bloom} J.~S.,  2006, \mn@doi [\araa]
  {10.1146/annurev.astro.43.072103.150558}, \href
  {http://adsabs.harvard.edu/abs/2006ARA%26A..44..507W} {44, 507}

\bibitem[\protect\citeauthoryear{{Woosley} \& {Heger}}{{Woosley} \&
  {Heger}}{2003}]{woosley_2003}
{Woosley} S.~E.,  {Heger} A.,  2003, ArXiv Astrophysics e-prints, \href
  {http://adsabs.harvard.edu/abs/2003astro.ph..9165W} {}

\bibitem[\protect\citeauthoryear{Woosley \& Heger}{Woosley \&
  Heger}{2006}]{woosley_2006a}
Woosley S.~E.,  Heger A.,  2006, \apj, 637, 914

\bibitem[\protect\citeauthoryear{{Yoon} \& {Langer}}{{Yoon} \&
  {Langer}}{2005}]{yoon_05}
{Yoon} S.~C.,  {Langer} N.,  2005, \mn@doi [\aap] {10.1051/0004-6361:20054030},
  \href {https://ui.adsabs.harvard.edu/abs/2005A&A...443..643Y} {443, 643}

\bibitem[\protect\citeauthoryear{{Yoon}, {Langer}  \& {Norman}}{{Yoon}
  et~al.}{2006}]{yoon_2006}
{Yoon} S.~C.,  {Langer} N.,   {Norman} C.,  2006, \mn@doi [\aap]
  {10.1051/0004-6361:20065912}, \href
  {https://ui.adsabs.harvard.edu/abs/2006A&A...460..199Y} {460, 199}

\bibitem[\protect\citeauthoryear{Yoon et~al.,}{Yoon et~al.}{2018}]{yoon_2018}
Yoon J.,  et~al., 2018, \mn@doi [The Astrophysical Journal]
  {10.3847/1538-4357/aaccea}, 861, 146

\makeatother
\end{thebibliography}



\bsp	
\label{lastpage}
\end{document}